\documentclass{article}

\usepackage{kr}

\usepackage{times}
\usepackage{soul}
\usepackage{url}
\usepackage[hidelinks]{hyperref}
\usepackage[utf8]{inputenc}
\usepackage[small]{caption}
\usepackage{graphicx}
\usepackage{amsmath}
\usepackage{amsthm}
\usepackage{amssymb}
\usepackage{booktabs}
\usepackage{algorithm}
\usepackage{algorithmic}
\usepackage{amsfonts}

\usepackage{textcomp}
\usepackage{subcaption}
\usepackage{array}
\usepackage{mathtools}
\usepackage{xspace}
\usepackage{multirow}

\newcommand{\ours}{\textsc{ASPen}\,}
\newcommand{\LACE}{\textsc{Lace}\,}

\newcommand{\tbl}[1]{Table~\ref{#1}}
\newcommand{\fig}[1]{Figure~\ref{#1}}
\newcommand{\sect}[1]{Section~\ref{#1}}
\def\EqRel{\mathsf{EqRel}}
\def\dcons{\mathsf{Dom}}
\newcommand{\link}{\mathsf{Eq}}

\def\ex{\mathsf{ex}}

\newcommand{\aspsong}{\texttt{song}}
\newcommand{\aspband}{\texttt{band}}
\newcommand{\aspappear}{\texttt{appear}}

\newcommand{\MaxSOL}{\mathsf{MaxSol}}
\newcommand{\SOL}{\mathsf{Sol}}

\newcommand{\E}{\Sigma}

\renewcommand{\S}{\mathcal{S}}

\newcommand{\ub}{\mathsf{UB}}
\newcommand{\lb}{\mathsf{LB}}
\newcommand{\MS}{\mathsf{MS}}
\newcommand{\PM}{\mathsf{PM}}
\newcommand{\magellan}{Magellan\,}
\newcommand{\jedai}{JedAI\,}
\newcommand{\vlog}{VLog4j\,}
\newcommand{\dblp}{\textit{Dblp}\,}
\newcommand{\cora}{\textit{Cora}\,}
\newcommand{\imdb}{\textit{Imdb}\,}
\newcommand{\music}{\textit{Mu}\,}
\newcommand{\cormusic}{\textit{MuC}\,}
\newcommand{\poke}{\textit{Poke}\,}

\newcommand{\dup}[1]{\mathsf{Du}#1}
\newcommand{\tground}{\mathbf{t_g}}
\newcommand{\tpground}{\mathbf{t_{pg}}}
\newcommand{\tolground}{\mathbf{t_{og}}}
\newcommand{\tsolve}{\mathbf{t_s}}
\newcommand{\tprep}{\mathbf{t_p}}
\newcommand{\ter}{\mathbf{t}_\text{o}}
\newcommand{\tero}{\mathbf{t}_\text{oo}}

\def\fone{\mathbf{F_1}}

\newcommand{\ground}[1]{\textit{gr}(#1)}

\newcommand{\ASet}[1]{\textit{SM}(#1)}
\newcommand{\aspenc}{\ensuremath{\Pi_{(D, \Sigma)}}\xspace}

\newcommand{\defneg}{\texttt{not}}
\newcommand{\linkasp}{\texttt{eq}\xspace}

\newcommand{\simpred}{\texttt{sim}\xspace}

\newcommand{\emptypred}{\texttt{empty}}

\def\ubpred{\texttt{ubeq}}

\newcommand{\enumi}{\textbf{i)}}
\newcommand{\enumii}{\textbf{ii)}}
\newcommand{\enumiii}{\textbf{iii)}}
\newcommand{\enumiv}{\textbf{iv)}}

\newcommand{\clingo}{\texttt{clingo}~} 
\newcommand{\python}{\texttt{Python}~}
\newcommand{\asprin}{\texttt{asprin}}
\newcommand{\xclingo}{\texttt{xclingo}~} 

\def\oursim{\mathsf{Sim}_\mathsf{opt}}
\def\oursimalg{sim_\mathsf{opt}}
\def\cssimalg{sim_\mathsf{cs}}

\newcommand{\titlep}[1]{ \smallskip \noindent \textbf{#1}}
\newcommand{\titleit}[1]{\smallskip \noindent \underline{\textit{#1.}}}

\usepackage{subfiles} 

\usepackage{tikz}
\usetikzlibrary{trees,automata,positioning}
\usepackage{enumitem}

\definecolor{darkgreen}{rgb}{0.0, 0.5, 0.0}
\definecolor{cadmiumgreen}{rgb}{0.0, 0.42, 0.24}
\definecolor{darkspringgreen}{rgb}{0.09, 0.45, 0.27}

\newcommand{\rj}[1]{\textcolor{red}{#1}}
\newcommand{\changed}[1]{\textcolor{darkspringgreen}{#1}}

\newtheorem{example}{Example}
\newtheorem{theorem}{Theorem}

\title{\textsc{ASPen}: ASP-Based System for Collective Entity Resolution}

\author{%
Zhiliang Xiang$^1$ \and  Meghyn Bienvenu$^{2,3}$ \and Gianluca Cima$^4$ \and\\ Víctor Gutiérrez-Basulto$^1$ \and Yazmín Ibáñez-García$^1$\\
\affiliations
$^1$Cardiff University, UK\\
$^2$Univ. Bordeaux, CNRS, Bordeaux INP, LaBRI, UMR 5800, Talence, France\\
$^3$Japanese-French Laboratory for Informatics, CNRS, NII, IRL 2537, Tokyo, Japan\\ 
$^4$Sapienza University of Rome, Italy\\
\emails
 \{xiangz6,gutierrezbasultov,ibanezgarciay\}@cardiff.ac.uk, 
meghyn.bienvenu@labri.fr,  cima@diag.uniroma1.it
}
\def\BibTeX{{\rm B\kern-.05em{\sc i\kern-.025em b}\kern-.08em
    T\kern-.1667em\lower.7ex\hbox{E}\kern-.125emX}}
\makeatletter

\newcommand{\thickhline}{%
    \noalign {\ifnum 0=`}\fi \hrule height 1pt
    \futurelet \reserved@a \@xhline
}

\newcolumntype{"}{@{\hskip\tabcolsep\vrule width 1pt\hskip\tabcolsep}}
\makeatother

\makeatletter
\newcommand{\thinhline}{%
    \noalign {\ifnum 0=`}\fi \hrule height 0.2pt
    \futurelet \reserved@a \@xhline
}
\newcolumntype{"}{@{\hskip\tabcolsep\vrule width 1pt\hskip\tabcolsep}}
\makeatother

\newcolumntype{?}{!{\vrule width 1.1pt}} 
\begin{document}

\maketitle

\begin{abstract}
In this paper, we present  \textsc{ASPen}, an answer set programming (ASP)
implementation of a recently proposed declarative framework 
for collective entity resolution (ER). While an ASP encoding had been
previously suggested, several practical issues had been neglected, 
most notably, the question of how to efficiently 
compute the (externally defined) similarity facts that are used in rule bodies. 
This leads us to propose new variants of the encodings (including Datalog approximations) 
and show how to 
employ different functionalities of ASP solvers to compute (maximal) 
solutions, and (approximations of) the sets of possible and certain merges. 
A comprehensive experimental evaluation of \textsc{ASPen}\ on real-world 
datasets shows that the approach is promising, 
achieving high accuracy in real-life ER scenarios. 
Our experiments also yield useful insights into the relative merits of 
different types of (approximate) ER solutions,
the impact of recursion, and factors influencing performance. 

\end{abstract}
\section{Introduction}
Entity resolution (ER) is a fundamental problem in data quality which aims at identifying different constants (of the same type) 
 that refer to the same real-world entity~\cite{DBLP:conf/kdd/GetoorM13,DBLP:conf/kdd/GetoorM13}. Over time, several variants of ER (also known as record linkage or deduplication) have been investigated, including pairwise matching in a single table and (the more general) \emph{collective} ER,
which looks at  the joint resolution (match, merge) of entity references across multiple tables~\cite{BhattacharyaTKDD07}.
 Given the multi-faceted nature of the ER problem, diverse techniques have been already proposed to tackle it~\cite{ChristophidesACMSurvey21}, including  machine  learning (ML), 
 and declarative frameworks based upon logical rules and constraints. 
Most existing approaches to ER {focus on single-pass matching of tuples within a single table or between a pair of tables}, 
and ML methods have obtained remarkable results~\cite{ditto-2020} for such settings.
On the other hand, declarative methods are well suited for complex multi-relational settings as they naturally exploit the relational dependencies to perform collective ER. Moreover, some declarative approaches conduct ER in a \emph{recursive} manner (also called \emph{deep} ER~\cite{DengICDE22}), instead of examining entity pairs only once. 
 %


\textsc{Lace} is a recently proposed declarative framework~\cite{lace_2022} for collective ER, which employs hard and soft rules to define mandatory and possible merges, and denial constraints~\cite{BertossiMC2011} to enforce consistency of the resulting database and constrain the allowed combinations of merges.  \textsc{Lace} employs a dynamic semantics in which rule bodies are evaluated over the current induced database, taking into account all previously derived merges. This makes it possible to support recursive scenarios, while ensuring that all merges have a (non-circular) derivation.  The semantics of  \textsc{Lace}  is also global  since all occurrences of the matched constants are merged, rather than only those constant occurrences used
in deriving the match. Additionally, \textsc{Lace} considers a space of maximal (w.r.t.\ set inclusion) solutions, which emerges from adopting denial constraints to enforce consistency  and restricting which merges can be performed together, effectively creating choices. From this, one can define the notions of   possible and certain merges, as those merges that belong to some, respectively all, maximal solutions.

While the theoretical foundations of \LACE have been already established, a real-life implementation is to-date absent.  
\citeauthor{lace_2022}  (\citeyear{lace_2022}) showed that \LACE solutions can be faithfully captured by {answer set programming (ASP)}~\cite{DBLP:books/sp/Lifschitz19,asp-in-prac-2012} stable models. 
Building upon this, in this paper we present \textbf{\textsc{ASPen}}, an ASP-based system   for collective  ER. \textsc{ASPen} deals with  several practical issues, including the question of how to efficiently 
compute the (externally defined) similarity facts that are used in rule bodies. 
It also implements new variants of the encodings (including Datalog approximations) 
and uses different functionalities of ASP solvers to compute (maximal) 
solutions, and (approximations of) the sets of possible and certain merges. 
A comprehensive experimental evaluation of \textsc{ASPen}\ on real-world 
datasets shows that the approach is promising, 
achieving high accuracy in real-life ER scenarios. 
Our experiments also yield useful insights into the relative merits of 
different types of (approximate) ER solutions,
the impact of recursion, and factors influencing performance, such as the degree of dirtiness and size of datasets.  \ours also leverages the xclingo~\cite{xclingo2} framework for explaining conclusions of ASP programs to compute the justification of a merge in a solution, making \ours a \emph{justifiable} framework for ER.
 \smallskip \noindent \textbf{Related Work} 
 We discuss prior work  on logic-based approaches to ER; for details on ML and probabilistic methods, see \cite{ChristophidesACMSurvey21}. 
 The \LACE~framework, underlying \textsc{ASPen},  shares some characteristics with other logic-based ER methods. Similar to approaches based on matching dependencies (MDs)~\cite{BertossiTCS13,DengICDE22}, \ours adopts a dynamic semantics, enabling recursive ER. 
 Like the Datalog-like approaches Dedupalog~\cite{ArasuICDE09} and Entity Linking (EL)~\cite{BurdickTODS2016} (and unlike MDs), \ours considers a global semantics. Finally, as in the EL framework, \ours does not consider only a single solution, but rather a space of maximal 
 solutions, leading to notions of 
 possible and certain merges.
 While ASP encodings have been proposed for MDs~\cite{BertossiTCS13,DBLP:conf/flairs/BahmaniB17}, no implementation nor evaluation are provided. Likewise,  
 there is no implementation of the EL framework. 
The closest existing implementations of logic-based ER are those of  Dedupalog~\cite{ArasuICDE09} and MRL~\cite{DengICDE22}. However,  they are not publicly available. 
The main distinguishing features of 
\ours compared to existing systems are as follows:  
\enumi~
Rather than computing a single (possibly non-optimal) solution, \ours is not only able to compute, with the guarantee of correctness, a space of maximal solutions but also approximations with different levels of granularity based on different reasoning modes.
\enumii~
To the best of our knowledge, \ours is the first system that is able to explicitly give justifications to merges. We finally note in passing that ASP-based approaches to database repair have been explored~\cite{EiterFGL08,MannaRT13,DBLP:conf/semweb/AhmetajDPS22}. 
Additionally, a bespoke ASP system for cleaning 
healthcare data has also been developed~\cite{DBLP:conf/lpnmr/TerracinaML13}.



All programs, code, experiments and data of \ours are  available at \url{https://github.com/zl-xiang/Aspen}. 

\def\consts{\mathbf{C}}
\def\certm{\mathsf{CM}}
\def\certmax{\mathsf{CM}_\mathsf{max}}
\def\certall{\mathsf{CM}_\forall}
\def\possmax{\mathsf{PM}_\mathsf{max}}
\def\possall{\mathsf{PM}_\forall}
\def\possm{\mathsf{PM}}
\def\lbm{\mathsf{LB}}
\def\ubm{\mathsf{UB}}
\def\np{\mathsf{NP}}
\def\conp{\mathsf{coNP}}
\def\qterms{\mathsf{terms}}

\def\simpreds{\mathcal{S}_{\mathsf{sim}}}

\section{Preliminaries}\label{sec:preli}
We assume infinite sets of \emph{constants} $\mathbf{C}$ and \emph{variables} $\mathbf{V}$. 
A \emph{(database) schema $\mathcal{S}$} consist of a finite set of relation symbols, each having an associated arity $k \in \mathbb{N}$.
We write $R/k \in \mathcal{S}$ to indicate that $R$ has arity $k$. 
A \emph{relational atom} (over schema $\mathcal{S}$) takes the form $R(t_1, \ldots, t_k)$ where $R/k \in \mathcal{S}$ and $t_i \in \mathbf{C} \cup \mathbf{V}$ for $1 \leq i \leq k$,
and we call $R(t_1, \ldots, t_k)$ a \emph{fact} if 
$\{t_1, \ldots, t_k\} \subseteq \mathbf{C}$. 
We say that $t_i$ \emph{occurs in position} $i$ of an atom $R(t_1, \ldots, t_k)$. 
A \emph{database} $D$ (over schema $\mathcal{S}$) is a finite set of facts (over $\mathcal{S}$). 
The set of constants occurring in a database $D$ is denoted by $\dcons(D)$. 
%
%
Sometimes 
it will prove more natural to 
employ \emph{attributes} rather than (unnamed) positions,  writing 
$R(A_1, \ldots, A_k)$  to indicate that $(A_1, \dots, A_k)$ are the attributes of $R$.

A \emph{conjunctive query (CQ)} over a schema $\mathcal{S}$ takes the form $q(\vec{x})= \exists \vec{y}.\varphi(\vec{x},\vec{y})$, 
where $\vec{x}$ and $\vec{y}$  are disjoint tuples of distinguished and quantified variables, 
and $\varphi(\vec{x}, \vec{y})$ is a conjunction of 
relational atoms over $\mathcal{S}$, with variables drawn from $\vec{x} \cup \vec{y}$. 
We use $\qterms(q)$ for the set of \emph{terms} of $q$, i.e.~the constants and variables appearing in~$q$. 
The set of \emph{answers} to a CQ $q(\vec{x})= \exists \vec{y}.\varphi(\vec{x},\vec{y})$ on a database $D$ (over the same schema), denoted $q(D)$,
contains those tuples of constants $\vec{c}$ such that there exists a mapping $h: \qterms(q) \rightarrow \dcons(D)$
such that (i) $h(\vec{x}) = \vec{c}$, (ii) $h(t)=t$ for $t \in \qterms(q) \cap \mathbf{C}$, and (iii) for every atom $R(t_1, \ldots, t_k)$ of $q$, 
$R(h(t_1), \ldots, h(t_k)) \in D$.
We will also consider \emph{CQs with inequalities} (CQ$^{\neq}$), which may 
additionally include inequality atoms $t_i \neq t_j$, in which case we require that the mapping $h$
further satisfies (iv) $h(t_i) \neq h(t_j)$ whenever $q$ contains $t_i \neq t_j$. 
When $q$ has only quantified variables, it is called \emph{Boolean}, 
and we say that a Boolean CQ$^{\neq}$ $q$ is \emph{satisfied} in $D$ if $q(D)=\{()\}$.
%
%
A \emph{denial constraint} (DC) takes the form 
$q \rightarrow \bot$, 
where 
$q$ is a Boolean CQ$^{\neq}$. We say that $D$ \emph{satisfies a DC} $q \rightarrow \bot$ just in the case that 
the CQ$^{\neq}$ $q$ is not satisfied in $D$. Functional dependencies (FDs) and primary key constraints
are special cases of DCs. 

When we speak of complexity, we will 
always mean \emph{data complexity}, which is
measured 
only in terms of the size of the input database, with all other inputs (e.g.\ 
ER specifications and queries) treated as fixed.


\section{Entity Resolution Framework}\label{sec:frame}
In this section, we recall the syntax and semantics of \textsc{Lace}~\cite{lace_2022}.
We 
also introduce 
new notions that are useful for our system. 

\subsection{\textsc{Lace} Entity Resolution Specifications}
Entity resolution can be formulated as the task of discovering pairs of syntactically distinct database constants that refer 
to the same entity (we will often use the term 
 \emph{merges} for such pairs). 
We adopt the
\textsc{Lace} framework, which 
focuses on identifying merges of entity-referencing constants (e.g.\ paper or author identifiers).
 
\textsc{Lace} 
employs hard and soft rules to identify mandatory and possible merges. 
\emph{Hard rules} and \emph{soft rules}, over schema $\mathcal{S}$, take respectively  the forms:
\begin{equation}\label{rulesform}
q(x,y) \Rightarrow \link(x,y), \quad  q(x,y) \dashrightarrow \link(x,y) 
\end{equation}
where $\link$ is a special relation symbol not in $\mathcal{S}$ used to store pairs of merged constants, and $q(x,y)$ is a CQ  
using relation symbols from  
$\mathcal{S}$. 
Intuitively, a hard (resp.\ soft) rule states that a pair of constants $(c_1,c_2)$ being an answer to $q$ provides sufficient (resp.\ reasonable) 
evidence that $c_1$ and $c_2$ refer to the same real-world entity. 
We use  $
q(x,y) \rightarrow \link(x,y) \in \Gamma$ to denote a generic (hard or soft) rule (note the arrow).  

\begin{figure*}[!htb]
    \begin{subtable}[t]{0.47\textwidth}
        \centering
        \caption*{Band(bid,~name,~genre,~year,~founder)}
        \begin{tabular}{| c | c | c | c | c|}
            \thinhline
            bid&name&genre&year&founder\\
            \thickhline
            $b_1$&Pink Floyd&Psy.~rock&1965&Barrett\\
            \thinhline
            $b_2$&The Pink Floyd&Prog.~rock&1965&Barrett\\
            \thinhline
        \end{tabular}
    \end{subtable}
    \hfill
    \begin{subtable}[t]{0.51\textwidth}
        \centering
        \caption*{Song(sid,~title,~lyricist,~bid)}
        \begin{tabular}{| c | c | c | c |}
            \thinhline
            sid&title&lyricist&bid\\
            \thickhline
            $s_1$&Shine On You Crazy Diamond (I-IV)&Waters&$b_1$\\
            \thinhline
            $s_2$&Shine On You Crazy Diamond&Waters&$b_2$\\
            \thinhline
            $s_3$&Shine On You Crazy Diamond (V-IX)&Waters&$b_1$\\
            \thinhline
        \end{tabular}
    \end{subtable}
    \begin{minipage}{0.586\textwidth}
        $\sigma_{\ex}\mathit{:}~\textit{Song}(x,t,l,b) \wedge \textit{Song}(y,t^\prime,l,b) \wedge t \approx t' \dashrightarrow \link(x,y)$\\
        $\rho_{\ex}\mathit{:}~\textit{Band}(x,n,g,d,f) \wedge \textit{Band}(y,n',g',d,f) \wedge n \approx n' \wedge g \approx g' \Rightarrow \link(x,y)$\\
        $\delta_{\ex}\mathit{:}~
        \textit{Appear}(s,a,i) \wedge \textit{Appear}(s,a,j) \wedge i \neq j 
        \rightarrow \bot$\\

        \noindent We assume that the extension of the similarity relation $\approx$ \\(restricted to $\dcons(D_\ex)$) is the reflexive and symmetric closure of \\$\{(n_1,n_2),(g_1,g_2),~(t_1,t_2),~(t_2,t_3)\}$, where $n_i$, $g_i$, and $t_i$ are \\the name and genre of band $b_i$ and the title of song $s_i$, respectively. 
    \end{minipage}
    \hfill
    \begin{minipage}{0.41\textwidth}
        \begin{subtable}[t]{\textwidth}
            \centering
            \caption*{Appear(sid,~album,~position)}
            \begin{tabular}{| c | c | c |}
                \thinhline
                sid&album&position\\
                \thickhline
                $s_1$&Wish You Were Here&$1$\\
                \thinhline
                $s_2$&A Delicate Sound of Thunder&$1$\\
                \thinhline
                $s_3$&Wish You Were Here&$5$\\
                \thinhline
            \end{tabular}
        \end{subtable}
    \end{minipage}
    \caption{A schema $\S_\ex$, an $\S_\ex$-database $D_\ex$, and an ER specification $\E_\ex=\langle \Gamma_\ex, \Delta_\ex \rangle$ for $\S_{\ex}$, where $\Gamma_\ex=\{\rho_\ex,\sigma_\ex\}$ and $\Delta_\ex=\{\delta_\ex\}$. 
    }\label{fig:ExLACE}
\end{figure*}

It is natural and useful for rule bodies to use \emph{similarity relations}, i.e.\ binary relations whose extension 
is fixed and computed using some external function (e.g.\ by applying a string similarity measure to a pair of 
constants and keeping those pairs whose score exceeds a given threshold). We shall thus allow the considered 
schema to contain such similarity relations and will adopt the more intuitive infix notation $x \approx y$ 
(possibly with indices) for similarity atoms.
Note also that in the present 
section, we will assume that the similarity facts are provided
as part of the input database,
leaving the question of how to best compute them to later sections. 
\begin{example}{\em
Figure~\ref{fig:ExLACE} presents our running example inspired by the MUSIC dataset.  
Relations $\textit{Group},~\textit{Song},~\textit{Appear}$ of the schema $\S_{\ex}$ 
provide information about music bands, songs, and which songs appear on which albums. The attributes \textbf{bid} and \textbf{sid} contain identifiers for bands and songs respectively, and the task at hand is to identify which of these identifiers refer to the same band / song. 
The hard rule $\rho_{\ex}$ says that if two band ids have the same founding year, same founder, similar names, and similar genres, they must refer to the same band. The soft rule $\sigma_{\ex}$ states that two song ids likely refer to the same song if they have the same band and lyricist and  similar titles. For simplicity, a single similarity relation $\approx$ is used to say which strings count as similar. 
}
\end{example}

To enforce consistency of the inferred merges and to help block false positives, 
\textsc{Lace} specifications may also include 
\emph{denial constraints}. For example, the denial constraint $\delta_{\ex}$ in Figure~\ref{fig:ExLACE} is an FD for $\textit{Appear}$, which forbids the occurrence of the same song in different positions within an album.

A  \textsc{Lace}  \emph{entity resolution (ER) specification} 
over schema $\mathcal{S}$ takes the form $\Sigma = \langle \Gamma,\Delta \rangle$, 
where $\Gamma = \Gamma_h \cup \Gamma_s$ is a finite set of hard and soft rules 
and $\Delta$ is a finite set of denial constraints, all over~$\mathcal{S}$. 
Rulesets are required to satisfy a \emph{sim-safety condition}, 
whereby the relation positions involved in merges must be distinct from those involved in similarity atoms
.  
The example specification $\E_\ex$ is sim-safe as the attributes involved in similarity atoms (\textbf{title}, \textbf{name}, \textbf{genre}) are different from those involved in merges (\textbf{bid}, \textbf{sid}). Thanks to sim-safety, \emph{we may assume w.l.o.g.\ that the 
constants appearing in merge positions do not occur in sim positions. }


\subsection{Semantics of \textsc{Lace}}
The semantics of \textsc{Lace} associates with every database $D$ and ER specification $\Sigma$
a set of solutions, where each solution takes the form of an equivalence relation over $\dcons(D)$, indicating 
which constants refer to the same entity. 
Given a set of pairs 
$P \subseteq S \times S$,
we write $\EqRel(P, S)$ to denote the least equivalence relation over $S$ that extends $P$. 


%
In a nutshell, ER solutions in \LACE are obtained by applying the hard and soft rules
in such a manner that all hard rules and constraints are satisfied, with the inferred 
$\link$-facts determining the equivalence relation. 
Importantly, 
the evaluation of \LACE rules takes into account previously derived merges, making it possible for merges to trigger further merges.
Formally, given a database $D$, and 
equivalence relation $E$ over $\dcons(D)$, the \emph{database induced by $D$ and $E$}, denoted $D_E$, 
is 
obtained from $D$ by replacing each constant $c$ by the (uniformly chosen) representative $\bar{c}$ of its equivalence class. 
%
The set $q(D,E)$ of \emph{answers to 
$q(\vec{x})$ w.r.t.\ $D$ and $E$} is defined as: 
$${q(D, E)= \{(c_1, \ldots, c_n) \mid (\bar{c_1}, \ldots, \bar{c_n}) \in q(D_E)\}}$$
Rule $
q(x,y) \rightarrow \link(x,y)$ 
is satisfied in $(D,E)$ if $q(D,E) \subseteq E$, and 
DC $\delta$ is satisfied in $(D,E)$ if $\delta$ is satisfied in $D_E$.



With these notions in hand, we can now define solutions. 
Given an ER specification $\Sigma$ and 
database $D$, 
we call $E$ a  \emph{candidate solution for $(D, \Sigma)$ } if 
one of the following holds: 
\begin{enumerate}
    \item $E = \EqRel(\emptyset, \dcons(D))$ \hfill 
   \item $E = \EqRel(E^\prime \cup \{\alpha \}, D)$, where $E^\prime$ is a candidate solution for $(D, \Sigma)$ and 
   $\alpha \in q(D,E) \setminus E'$ for some $q(x,y) \rightarrow \link(x,y) \in \Gamma$ 
\end{enumerate}
and it is a \emph{solution} for $(D, \Sigma)$ if additionally (i)~$(D, E) \models \Gamma_h$ and (ii)~$(D, E) \models \Delta$. 
We denote by $Sol(D, \Sigma)$ the set of solutions for $(D, \Sigma)$,
and let $\MaxSOL(D, \Sigma)$ be the set of \emph{maximal solutions}, i.e. solutions $E$ such that there is no solution $E^\prime$ for $(D, \Sigma)$ with $E \subsetneq E^\prime$. 

\begin{example}
{\em    We determine the maximal solutions for $(D_{\ex},\Sigma_\ex)$. The initial trivial equivalence relation $E = \EqRel(\emptyset, \dcons(D_{\ex}))$ is not a solution 
    as the hard rule $\rho_{\ex}$ requires us to merge $b_1$ and $b_2$. 
    Applying $\rho_{\ex}$, we obtain $E' = \EqRel(\{(b_1,b_2)\}, \dcons(D_{\ex}))$, which is a solution for $(D_{\ex},\Sigma_\ex)$ but not a maximal one. Indeed, due to the addition of $(b_1,b_2)$, both $(s_1,s_2)$ and $(s_2,s_3)$ can now be obtained by applying the soft rule $\sigma_{\ex}$.
    Notice, however, that it is not possible to include both of them, 
    as transitivity would force us to include $(s_1,s_3)$, 
    leading to a violation of $\delta_{\ex}$. 
    We thus obtain two maximal solutions for $(D_{\ex},\Sigma_\ex)$, namely $E_1 = \EqRel(\{(b_1,b_2),(s_1,s_2)\}, \dcons(D))$ and $E_2 = \EqRel(\{(b_1,b_2),(s_2,s_3)\}, \dcons(D))$.}
\end{example}

There can be zero, one, or multiple (maximal) solutions.
However, for specifications that do not contain soft rules, there can be at most one solution, 
and for specifications that do not contain constraints, there is always a single solution. 

\subsection{Summarizing and Explaining Solutions} \label{merge-explain}

When there are multiples solutions, it is useful to be able to summarize them
by identifying merges that occur in all or some maximal solutions. 
Formally, we say that a merge $\alpha$ is \emph{certain} if 
$\alpha \in E$ for every $E \in \MaxSOL(D, \Sigma)$,
and it is \emph{possible} if $\alpha \in E$ for some $E \in \MaxSOL(D, \Sigma)$ (equivalently, some $E \in \SOL(D, \Sigma)$).
We use $\certm(D,\Sigma)$ and $\possm(D,\Sigma)$ 
for the sets of certain and possible merges.

While certain and possible merges provide natural summarizations, 
they are unfortunately hard to compute: 

\begin{theorem}\cite{lace_2022}
It is $\np$-complete (resp.\ $\Pi^p_2$-complete) in data complexity to decide if $\alpha \in \possm(D,\Sigma)$ (resp.\ $\certm(D,\Sigma)$).
\end{theorem}

For this reason, it will prove useful to define efficiently computable approximations.  
To this end, we consider two ways of simplifying a specification $\Sigma=\langle \Gamma_h \cup \Gamma_s,\Delta \rangle$:
\begin{itemize}
\item $\Sigma^{\mathsf{lb}}= \langle \Gamma_h, \emptyset \rangle$, i.e.\ remove soft rules and constraints
\item $\Sigma^{\mathsf{ub}}= \langle \Gamma_h \cup \Gamma_{s\rightarrow h}, \emptyset \rangle$, i.e.\ 
drop constraints and replace soft rules by the corresponding hard rules ($\Gamma_{s\rightarrow h}$)
\end{itemize}
As $\Sigma^{\mathsf{lb}}$ and $\Sigma^{\mathsf{ub}}$
do not contain any constraints, they will always yield a unique solution. We can
therefore define 
\begin{itemize} 
\item $\lbm(D,\Sigma)$ as the unique solution to $(D, \Sigma^{\mathsf{lb}})$
\item  $\ubm(D,\Sigma)$ as the unique solution to $(D, \Sigma^{\mathsf{ub}})$ 
\end{itemize}
We summarize the properties of the different merge sets: 

\begin{theorem}
The sets $\lbm(D,\Sigma)$ and $\ubm(D,\Sigma)$ can be computed in polynomial w.r.t.\ data complexity.
Moreover, if $Sol(D, \Sigma) \neq \emptyset$, then 
$$\lbm(D,\Sigma)\subseteq \certm(D,\Sigma) \subseteq \possm(D,\Sigma)\subseteq \ubm(D,\Sigma)$$
and $\certm(D,\Sigma) \subseteq M \subseteq \possm(D,\Sigma)$ for $M \! \in \MaxSOL(D, \Sigma)$.
\end{theorem}

\begin{example}{\em
In our toy example, $\lbm(D_{\ex},\Sigma_{\ex})$ and $\certm(D_{\ex},\Sigma_{\ex})$ 
coincide, and the only non-trivial pair they contain is $(b_1,b_2)$. 
The set $\possm(D_{\ex},\Sigma_{\ex})$ further contains merges $(s_1,s_2)$ and $(s_2,s_3)$, 
whilst $\ubm(D_{\ex},\Sigma_{\ex})$ 
also 
contains 
$(s_1,s_3)$ (not present in any solution). }
\end{example}

We will employ \emph{proof trees} to explain why a merge appears in a solution (we provide here some intuitions behind proof trees and refer to App. \ref{sec:exp-app} for more  details). Informally, a proof tree for a merge $\alpha$ in a solution $E \in Sol(D, \Sigma)$ is a node-labelled tree such that (a) the root node has label~$\alpha$, 
(b) every leaf node is labelled with a fact from $D$,
and (c) every non-leaf node~$n$ is labelled with a pair of constants $(d,e)$ corresponding to a single transitive step or a rule application. 

To quantify the number of successive rule applications needed to obtain a merge, we let the \emph{rule-depth of a proof tree} $T$ be the maximum number of rule nodes in any leaf-to-root path in $T$. The \emph{level of a merge $\alpha$ in a solution $E$} is $0$ if $\alpha=(c,c)$ for some $c\in \consts$, and otherwise is the minimum rule-depth of all proof trees of $\alpha$ in $E$. 

\section{ASP Encoding and Algorithms}\label{sec:asp}
The \ours system implements the \LACE framework 
by   
encoding ER specifications as ASP programs and making calls to ASP solvers
to generate and reason about ER solutions. 
After recalling some ASP notions,
we present the ASP encoding and algorithms employed by \textsc{ASPen}. 
We also explore some practical issues that were ignored in the theoretical treatment of \LACE, most importantly, the question of how to handle similarity atoms. 

%
We  assume familiarity with ASP basics, see~\cite{DBLP:journals/cacm/BrewkaET11,asp-in-prac-2012,DBLP:books/sp/Lifschitz19} for more details. For our purposes, it is enough to consider \emph{normal rules} and \emph{constraints}. That is, respectively, rules with a single atom head and rules with an empty head. We shall use $\Pi$ to denote an ASP \emph{program} (a set of rules). Given a program $\Pi$, we use $\ground{\Pi}$ to denote the set of all ground instantiations  rules from $\Pi$ with constants occurring in $\Pi$, and  
$\ASet{\Pi}$ to denote the set of all stable models of~$\Pi$. 
Determining whether a program has a stable model is the fundamental decision problem in ASP, which is solved using ASP solvers~\cite{DBLP:conf/ijcai/GebserLMPRS18}. However, more reasoning modes are needed to cover problems encountered in practice. Most modern ASP solvers are also able to \emph{enumerate} ($n$ elements of) $\ASet{\Pi}$; \emph{project} answers w.r.t.\ a given set of atoms, and enumerating ($n$ elements of) those projections; computing the intersection (resp. union) of all stable models of $\Pi$ (\emph{cautious}, resp. \emph{brave} reasoning); and perform \emph{optimisations} by  computing some (or enumerating $n$) elements of $\ASet{\Pi}$ that minimize a given objective function. 
\subsection{ASP Encoding of Solutions}\label{sec:asp-enc}
Given a \LACE specification $\Sigma$ and a database $D$, we define an ASP program  \aspenc 
containing all the facts in $D$, and an ASP rule for each (hard or soft) rule in $\E$.  
Consider, for example,  the specification $\E_\ex$ in Figure~\ref{fig:ExLACE}. Rules $\mathbf{\rho_{\ex}}$, $\sigma_{\ex}$ and  $\delta_\ex$ are translated as follows: 
\begin{small}
\[
       \begin{aligned}
      \linkasp(X,Y) &\leftarrow  \aspband(X,N,G,D,F), \aspband(Y,N^\prime,G^\prime,D,F), 
        \\
        &\simpred(G,G^\prime,S), S\geq 95, \simpred(N,N^\prime,S^\prime), S^\prime \geq 95. \\[5pt]
     \{\linkasp(X,Y)\} &\leftarrow \aspsong(X,T,L,B), \aspsong(Y,T^\prime,L,B^\prime), 
        \\
        & \defneg~\emptypred(L), \simpred(T,T^\prime, S), S\geq 95, \linkasp(B,B^\prime). \\
       \end{aligned}
       \]
   \[ 
      \bot \leftarrow \aspappear(S,A,I), \aspappear(S^\prime,A,J), \linkasp(S,S^\prime), I\not = J.
         \]
\end{small}
Roughly, each relational atom in the body of a (hard or soft rule) is translated into an atom in the body of an ASP rule. The relation $\linkasp$ is used to store mandatory merges (and ultimately solutions to $(D, \Sigma)$). Atoms of the form $\linkasp(X,X')$ in the rule bodies are used to encode that instantiations of $X$ and $X'$  have been determined to denote the same entity.  \aspenc also includes rules for encoding that $\linkasp$ is an equivalence relation. 
Soft rules are encoded using a choice rule encoding the possibility of including $(X,Y)$ in $\linkasp$. We note that choice rules are special syntactic sugar available in ASP. 

Differently from the original ASP encoding in ~\cite{lace_2022}, we encode similarity relations with a relation $\simpred_i(X,Y,S)$, where the first two arguments store the pair of constants to be assessed for similarity, while the third argument corresponds to a similarity score. This offers the flexibility of tuning the threshold for the same similarity measure.    Facts of the form $\simpred_i(X,Y,S)$ are included in $D$ after a preprocessing stage, as discussed in Section~\ref{sec:sim}. 
Notably, the encoding requires a special treatment of null values. 
Missing values in databases might be problematic~\cite{2012Fan-book}, in particular, when joins need to be performed to evaluate rule bodies. 
To represent null values in $\Pi_{(D, \Sigma)}$, we use an atom $\textit{\emptypred}(nan)$, where $nan$ is a special constant.  To encode the fact that merges are not performed on unknown or missing values,  we add an atom of the form $\textit{\defneg~\emptypred}(V)$ in rule bodies, for every joined variable $V$. This encoding  prevents merges that otherwise would result from rule bodies being satisfied when considering two nulls equivalent. With the encoding \aspenc in place, solutions of $(D,\Sigma)$ are then obtained by projecting stable models of $\aspenc$  w.r.t.\  $\linkasp$.   %
\begin{theorem}[\cite{lace_2022}] \label{thm:correct}
For every database $D$ and ER specification $\Sigma$: 
$E \in Sol(D, \Sigma)$  iff $E = \{(a,b) \mid \linkasp(a,b) \in M)\}$ 
for some  stable model $M \in \ASet{\aspenc}$. 
In particular, $Sol(D,\Sigma) \neq \emptyset$ iff $\ASet{\aspenc} \neq \emptyset$\footnote{We note that the proof in \cite{lace_2022} does not consider nulls, but it can be easily extended.}.
\end{theorem}
%

\subsection{ASP-based Algorithms}\label{sec:asp-alg}
We  now explain how to employ the ASP encoding to generate (maximal) solutions 
and other sets of merges, cf.\ Sec.\ref{merge-explain}.


\titleit{Solutions} Thanks to Theorem \ref{thm:correct}, we can obtain a single solution from $Sol(D, \Sigma)$
by using the ASP solver to generate a stable model of $\aspenc$, then projecting onto its $\linkasp$  facts. 
Likewise, we can enumerate all or a fixed number of 
solutions by requesting an enumeration of
$\ASet{\aspenc}$.

\titleit{Maximal solutions} The maximal solutions correspond to the stable models of $\aspenc$ having a $\subseteq$-maximal set of $\linkasp$ facts. 
We rely on \asprin, a framework for implementing preferences among the stable models of a  program~\cite{asprin-2015}. 
In our case, we prefer a model $M'$ over $M$ if its projection to $\linkasp$ is a proper superset of that of $M$.
\asprin~also allows us to compute $n$ optimal stable models of a program.


\titleit{Lower and upper bound merge sets} To compute $\lbm(D,\Sigma)$, we first construct the ASP encoding $\Pi_{(D,\Sigma^{\mathsf{lb}})}$ based on $\Sigma^{\mathsf{lb}}$ and then we 
use the ASP solver to compute the (unique) answer set of $\Pi_{(D, \Sigma^{\mathsf{lb}})}$, which we project onto the $\linkasp$ relation. We proceed analogously for $\ubm(D,\Sigma)$, but using $\Sigma^{\mathsf{ub}}$. 


$
\begin{aligned}
         & \linkasp(X,Y) \leftarrow\aspsong(X,T,L,B), \aspsong(Y,T^\prime,L,B^\prime), 
        \\
        & \defneg~\emptypred(L), \simpred(T,T^\prime, S), S\geq 95, \linkasp(B,B^\prime).
\end{aligned}
$


\titleit{Possible merges} To generate $\possm(D,\Sigma)$, it suffices to run the ASP solver in brave reasoning mode,
and to project onto the $\linkasp$ relation. 
To check whether a particular pair $(c,c')$ is a possible merge, 
we can run the solver on $\aspenc \cup \{ \bot \leftarrow \text{not } \linkasp(c,c^\prime)\}$
observing that $(c,c') \in \possm(D,\Sigma)$ iff this modified program admits a stable model.


\titleit{Levels}
To support our analysis of the impact of recursion, we will need a means of
retrieving, for a given solution $E \in Sol(D, \Sigma)$, all triples 
$(c,c',i)$, where $\alpha=(c,c') \in E$ and $i$ is the level of $\alpha$ in $E$. This can be achieved by considering a variant $\Pi^{\mathsf{lvl}}(i)$ of $\Pi_{(D,\Sigma^{\mathsf{ub}})}$, which is an ASP program that takes an integer $i$ as a parameter and, by taking into consideration also the already derived merges inside $E$ (i.e.~those with level $<i$), applies single transitive steps or rule applications, attaching integer $i$ as the additional third element of the merges obtained in this way. 

\subsection{Similarity Computation}\label{sec:sim}
\def\truesim{\mathsf{Sim}_\mathsf{all}}
\def\oursim{\mathsf{Sim}_\mathsf{opt}}
\def\simfn{f_\approx}
\def\getsim{\texttt{getsim}}
\def\ubpred{\texttt{ubeq}}
Our ASP rules contain body atoms 
$\simpred_i(X,X^\prime, S)$, 
but such similarity facts are not present in the data sources 
and need to be computed via external functions. The question then is how best
to compute a sufficient set of $\simpred_i$ facts to properly evaluate the program, 
while avoiding making calls to the external functions for every possible pair of values. 


A naïve approach would be to compute and store
the set $\truesim(D)$, containing all facts $\simpred_i(c,d,s)$ 
such that $c$ and $d$ are data values of a form compatible with $\simpred_i$
and $s=f_i(c,d)$, with $f_i$ the function underlying the similarity relation $\simpred_i$. 
Although it requires only a polynomial number of function calls w.r.t.\ $|D|$,
it is nevertheless extremely costly on even moderately-sized databases. 
A first improvement would be to only consider those pairs of constants $(c,d) \in R[i] \times R'[j]$\footnote{$R[i]$ denotes the constant at position $i$ of $R$.}
such that there is a rule $\rho$ which contains an atom $\simpred_i(X,X^\prime, S)$
in which $X$ and $X'$ appear respectively in the $i$th (resp.\ $j$th) position of an $R$-atom (resp.\ $R'$-atom). However, as our experiments will show, this improved approach remains memory-consuming and its time consumption grows as the data size grows. 

Another idea would be to exploit the structure of the rules so that we 
only call the similarity functions on pairs of constants that occur in 
a similarity atom for which the rest of the rule body is satisfied. 
For example, for rule $ \sigma_{\ex}$, we would remove 
$\simpred(T,T^\prime, S)$ 
and instead store the compared variables ($T,T^\prime$) in a fresh relation $\getsim$ as follows:  
 \begin{align} \notag
 \getsim(T,T^\prime) \leftarrow &\aspsong(X,T,L,B), \aspsong(Y,T^\prime,L,B^\prime),  
        \\ 
        & \defneg~\emptypred(L), \linkasp(B,B^\prime). \label{modrule}
\end{align}
Note however that the body still 
contains $ \linkasp(B,B^\prime)$, whose extension is initially unknown.
We shall therefore employ a further ingredient: an overapproximation of the $\linkasp$ facts.
For this, we could use the program $\Pi_{(D,\Sigma^{\mathsf{ub}})}$, except that we do not know how to 
evaluate similarity atoms in rule bodies. 
One option would be to weaken the bodies by dropping all similarity atoms, but this 
yields a very loose approximation (cf.\ Appendix \ref{app:lam}).  
An alternative is to run the original $\Pi_{(D,\Sigma^\mathsf{ub})}$ program, 
but making \emph{online calls to the similarity functions}\footnote{Modern ASP solvers, e.g. \texttt{clingo}, support the syntax of external function calls~\cite{h2b-asp-2023}} by replacing literals $\simpred(X,Y,S), S\geq \delta$ with  $\texttt{sim}_\mathsf{ext}(X,Y)\geq\delta$. We denote by $\Pi_{(D,\Sigma^\mathsf{ub}_\mathsf{ext})}$ the modified program.
This will not only 
give us an 
upper bound on the true set of 
$\linkasp$ facts, but it will also compute a portion of the similarity facts.
 {In fact, a further alternative would be to rely entirely on an online computation of similarity atoms over the original program. However,
this is less efficient as results cannot be reused to compute different merge sets and may be time-consuming for the computation of $\MaxSOL(D,\Sigma)$ and $\PM(D,\Sigma)$ (we refer to Appendix~\ref{app:lam} for details).} 



Combining these ideas, we obtain the following 
approach to similarity computation, on input $D, \Sigma$. 
\begin{description}
\item[Phase 1] Compute the unique stable model $M_\mathsf{ub}$ of $\Pi_{(D,\Sigma^{\mathsf{ub}}_\mathsf{ext})}$
using 
an ASP solver with external function calls enabled. 
Let $U= \{\ubpred(c,d) \mid \linkasp(c,d) \in M_\mathsf{ub}\}$,
and let $L$ contain all $\simpred_i$ facts produced during the computation. 
\item[Phase 2] Let $\Pi_\simpred$ contain, for each rule $\tau$ in $\Pi_{(D, \Sigma)}$
that encodes a (soft or hard) rule in $\Sigma$ with at least one similarity atom in its body, 
a rule $\tau_\simpred$ that is obtained from $\tau$ by (i)~deleting all similarity atoms and all associated comparison atoms {(e.g. $ S\geq \delta$)}, 
(ii)~changing the head atom to $\getsim_i(T,T')$, where $\simpred_i(T,T^\prime, S)$ occurs in $\tau$,
and (iii)~renaming $\linkasp$ as $\ubpred$. 
 
\item[Phase 3] 
Let $M_\simpred$ be the unique stable model of $\Pi_\simpred \cup D \cup U$.
For every pair $(c,d)$ such that 
$\getsim_i(c,d) \in M_\simpred$
and there is no $\simpred_i(c,d,\_)$ fact in $L$, call 
the associated similarity function $f_i$ on input $(c,d)$,
and create the fact $\simpred_i(c,d,f_i(c,d))$. 
Let $\oursim(D,\Sigma)$ contain $L$
and all of these newly created similarity facts. 
\end{description}
%
Note that in Phase 2, our example rule $ \sigma_{\ex}$
would be replaced by \eqref{modrule}, but with 
$\ubpred(B,B^\prime)$ 
in place of $\linkasp(B,B^\prime)$. 

The next 
result shows 
that this 
approach 
is correct, i.e. we get the same $\linkasp$ facts using $\oursim$ 
rather than the full $\truesim$. 

\begin{theorem}\label{th:simCorrect}
For every database $D$ and ER specification $\Sigma$,
\[\{M_\linkasp \mid M \in \ASet{\Pi_{(D_\mathsf{all}, \Sigma)}} \} = \{M_\linkasp \mid M \in \ASet{\Pi_{(D_\mathsf{opt}, \Sigma)}} \} \]
where $D_\mathsf{all}= D \cup \truesim(D)$,  $D_\mathsf{opt} = D \cup \oursim(D,\Sigma)$, 
and $M_\linkasp$ is the set of $\linkasp$\, facts in $M$. 
\end{theorem}

\section{The \ours System}\label{sec:system}
\begin{figure}[!htb] 
    \centering
    \includegraphics[width=0.48\textwidth]{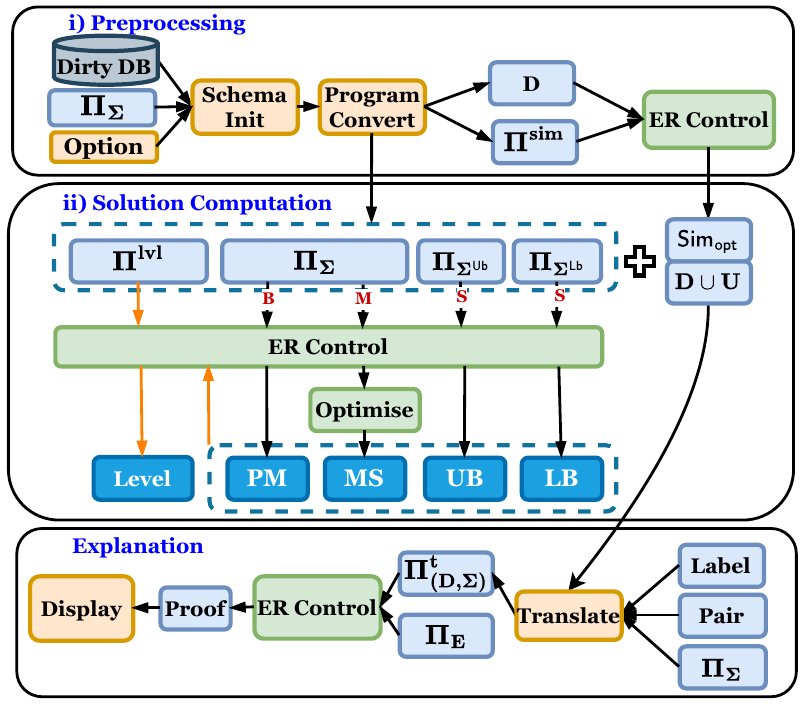}
    \caption{A General Pipeline of \ours.
    }
    \label{fig:pip_new}
\end{figure}
An overview of the pipeline of \ours is shown in Figure~\ref{fig:pip_new}, where boxes coloured light blue, strong blue, orange and green indicate an ASP program, ER solution, \python~program and \texttt{Python}-based ASP solver, respectively.  Arrows interconnecting these boxes symbolize the flow of data. The blue dashed boxes highlight a pool of candidates that may be chosen based on the input options. 

\ours takes as input {an ASP encoding of an ER specification (cf.\ Sec.\ \ref{sec:asp-enc})}, a set of running options (see below) and a dataset {in \texttt{CSV/TSV} format} containing duplicates with corrupted values, such as missing ones. {Depending on the input options, \ours~outputs either the specified merge sets as shown by the deep blue boxes in the second row of Figure  \ref{fig:pip_new}, and additionally explanations of all/particular merges (cf.\ Appendix \ref{sec:exp-app})}.
%
\ours comprises two primary phases for generating (approximate) ER solutions:
\begin{itemize}[itemsep=0.3ex, leftmargin=5mm]
 \item [\enumi] \textit{Preprocessing Phase:}  \ours initialises schema information and converts DB tuples into ASP facts, then computes similarity facts.
\item[\enumii] \textit{Solution Computation Phase:}  \ours  takes the DB and similarity facts computed before as input, then it computes various types of \LACE-based ER  solutions as specified in the input options.
\end{itemize}
The \emph{explanation phase} (third row, Fig.~\ref{fig:pip_new}) additionally takes as input a 
set of rule labels and a 
merge pair to check, and  visually displays proof trees of explanations 
(see App.~\ref{sec:exp-app}). 



\titleit{Preprocessing} 
{Initially, \ours uses a \python function called the \emph{Schema Initialiser} to load schema information from the input, including database tuples, schema names, relations, attributes, and foreign keys. This information is stored in a schema instance (\python object), which is used throughout the system workflow.
The schema information is then passed to
the \emph{Program Converter (PC)}, which generates variants of $\Pi_\Sigma$ and facts for the computation of different merge sets
~(cf.\ Sec.~\ref{sec:asp-alg}).
During preprocessing, the PC utilises the relation and attributes structure in the schema instance to convert the database tuples into facts, including the reflexive closure of constants on merge positions.  Additionally, empty entries in facts are replaced with the special constant $nan$.
To precompute the similarity facts, the PC transforms the input ER program into similarity filtering programs $\Pi^{\text{sim}}$. The resulting programs, along with the generated database facts, are 
passed to the \emph{ER Controller (ERC)}, a \python object that encapsulates reasoning facilities of the ASP solver \clingo\cite{asp-in-prac-2012} and algorithms (cf.\ Sec.~\ref{sec:asp-alg}). The ERC then executes the similarity filtering algorithm.
The resulting set of triples (compared pair of constants and score)
are stored as ternary \simpred-facts, which are then combined with database facts and the set of $\ubpred$s to form the fact base, as indicated by the flow arrow from the first row to the second row in \fig{fig:pip_new}. }

\titleit{Computing solutions}
With the fact base as input, the ERC selects a program variant  from the PC based on the selected solution option, as shown in the blue dashed box. 
Subsequently, the ERC executes grounding and 
solving with  one of the corresponding reasoning modes, as indicated by the red initials in~\fig{fig:pip_new}. 
It computes  the desired solutions as follows: 
\begin{enumerate}[itemsep=0.5ex, leftmargin=5mm]
    \item for the lower and upper bound merge sets ($\lb$, $\ub$)  
    by making a \textbf{S}tandard grounding\&solving call to \clingo.
    \item for a set of maximal solutions $\MS$-$n$ with an enumeration limit of $n$, with standard grounding calls on $\Pi_\Sigma$ but employing \textbf{M}aximisation solving calls supported by the optimiser~\asprin.
    \item for a set of all possible merges $\PM$ using the \textbf{B}rave reasoning mode of \clingo.
\end{enumerate}
In addition, if the merge level is required, the derived merge set will be further combined with the level retrieval program $\Pi^{\mathsf{lvl}}$ to assign levels to each of the merges, 
cf.\ orange arrows in the second row of~\fig{fig:pip_new}.

\titleit{Explanation}
{\ours~uses several \xclingo functions~\cite{xclingo,xclingo2} to provide explanations of merges, including: \enumi~the \texttt{Translate} function, which takes as input a program $\Pi$ and  labels declared upon $\Pi$, and outputs a program $\Pi^t$, 
including auxiliary rules to track the original rules that have been triggered, and  \enumii~the explanation program $\Pi_E$, which constructs explanations by grounding and solving together a stable model $M$ of $\Pi$ and $\Pi^t$.}

{As illustrated by the third row of~\fig{fig:pip_new},  \ours takes as input a merge pair to be justified and optionally a set of rule labels. A default set of labels is generated automatically capturing rule bodies of the ER program, if no rule labels are provided. Utilising the \texttt{Translate} function,
\ours translates the rules and fact base into a trace program $\Pi^t_{(D,\Sigma)}$ according to the labels. The ERC then combines this with the fact base and the \xclingo explanation program to derive a proof answer set, which is displayed as a graphical proof tree, explaining the input merge.
} 
\section{Experiments}\label{sec:experiments}
We conducted experiments on real-life ER scenarios, including both pairwise and complex multi-relational datasets. We evaluate the following aspects: (1) effectiveness of  our approach to similarity computation, (2) accuracy and efficiency  of \textsc{ASPen}, (3)  effect of recursion,  (4) factors impacting scalability and (5) use of Datalog engines to compute $\ub$ and $\lb$ approximations. We also exemplify generated justifications of merges in Appendix~\ref{sec:exp-app}.
\subsection{Experimental Setup}
\titleit{Datasets}
We  consider two  pairwise matching datasets from the bibliographic domain:  \textsc{DBLP-ACM}~\cite{DBLP-10} and   \textsc{CORA}, and 
four multi-relational datasets: 
\enumi~a subset of the  IMDB movie dataset~\cite{DengICDE22};
\enumii~a Music dataset
sampled from the Musicbrainz database with synthetic duplicates; 
\enumiii~ a dirtier instance of the Music dataset, which contains the same amount of duplicates 
but a higher percentage of nulls and more syntactical variants on the duplicates;
\enumiv~a Pokémon dataset,  
sampled from the Pokémon database with synthetic duplicates and complex inter-table references. 
For simplicity, we refer to the datasets as \textit{Dblp}, \textit{Cora}, \textit{Imdb}, \textit{Mu}, \textit{MuC} and \textit{Poke}, respectively. Sources and statistics of the datasets can be found in Appendix \ref{sec:setup-app}, as well as details about duplicates generation, sampling of datasets and the experimental environment.

\titleit{Similarity measures and metrics}
We calculate the syntactic similarities of constants based on their data types~\cite{erblox-2017}.   We adopt  the commonly used metrics~\cite{DBLP-10} of \emph{Precision} (\textbf{P}), \emph{Recall} (\textbf{R}) and \emph{F1-Score} ($\fone$). See App.~\ref{sec:setup-app} for  details.
\subsection{Similarity Filtering}~\label{sec:sim_block}
We denote by $\oursimalg$ the similarity algorithm presented in~\sect{sec:sim}. We implemented and ran $\oursimalg$  on all datasets and compared with the baseline procedure $\cssimalg$, which 
computes the similarity of every possible pair of values~\cite{erblox-2017}
occurring in every pair of sim positions appearing in rule bodies. 
\begin{table}[!htp]
\renewcommand\arraystretch{0.3}
\setlength{\tabcolsep}{0.2em}
\centering
\begin{tabular*}{\linewidth}{@{}ccc|cc|ccc@{}}
\toprule
\textbf{Data} & \textbf{\#At}  & \textbf{\#Cat} & $\mathbf{t}_{\mathsf{cs}}$ &$\mathbf{t_{\mathsf{opt}}}$ & $\textbf{M}_{\mathsf{cs}}$ &  $\textbf{M}_{\mathsf{opt}}$  & \textbf{MRed.}\\
\midrule
\dblp &  5& 10.2M & \textbf{96.3} & 531.4 & 512Mb &256Mb  & 50\\
\midrule
\cora &  17 & 0.8M &  \textbf{5.49} &932.9 & 32Mb &4Mb  & 87.5\\
\midrule
\imdb& 22&  19.6M  & \textbf{89.5} &  598.9  & 512Mb &128Mb  &75\\
\midrule
\music & 72&143.6M  &  \textbf{664.03} &772.3 & 4Gb &256Mb  & 93.8\\
\midrule
\cormusic & 72& 147.1M  &  867.5 & \textbf{446.4} & 4Gb &256Mb& 93.8\\
\midrule
\poke & 104& 769.9M&  9,419 & \textbf{4,142} & 32Gb & 128Mb & 99.6\\
\bottomrule
\end{tabular*}
\caption{
\#-columns denote the number of attributes, the sum of cross-products of constant pairs found in sim positions of a dataset. \textbf{MRed.} stands for the reduction rates of memory usage.
}
\label{tbl:sim-alg}
\end{table}
 
\titleit{Results}
 \tbl{tbl:sim-alg} presents the results on execution time ($\textbf{t}_{\{\mathsf{opt},\mathsf{cs}\}}$) and memory usage ($\textbf{M}_{\{\mathsf{opt},\mathsf{cs}\}}$).
We  observe that $\oursimalg$ requires substantially less space than $\cssimalg$, with memory usage reduction rates of   50\%, 87.5\%, 75\% and 99.6\%  on \textit{Dblp}, \textit{Cora}, \textit{Imdb} and \textit{Poke}, and  93.8\% on \music/\textit{MuC}, respectively.
Note that in general the reductions become substantially larger  as the number of attributes  increases from 5 to 104. This is because a key strength of $\oursimalg$ is to leverage joins on attributes as preconditions for two constants to be compared. Consequently, as the number of attributes within a schema increases, a proportional increase occurs in the preconditions that can be employed to restrict unnecessary comparisons. 

As for running time, $\cssimalg$ tends to be  faster than $\oursimalg$ in datasets of smaller scale, yielding speed advantages of 5.5, 6.5 and 100+ times on \textit{Dblp}, \textit{Imdb} and \textit{Cora}, respectively.  
Note  that as \textit{\#Cat} increases this tendency is inversed: on  \music both have similar running times,  and then on \cormusic and \poke,  $\oursimalg$ becomes 2 times faster. This shift might be because time spent on similarity computations  outweighs the $\oursimalg$ program execution time as the number of pairs to be compared grows quadratically for $\cssimalg$. 
\begin{table}[!htp]
\renewcommand\arraystretch{0.3}
\setlength{\tabcolsep}{0.25em}
\centering
\begin{tabular*}{\linewidth}{@{}cc|ccc|ccc@{}}
\toprule
\textbf{Data} & \textbf{Method}  & \multicolumn{3}{c|}{$\fone$~~~~~(\textbf{P}~~  
 /~~~\textbf{R})} & \textbf{$\ter$} & $\tground$ & $\tsolve$\\
\midrule
\multirow{6}{*}[-2.5ex]{\rotatebox{90}{\dblp}} 
& Magellan & 79.97& 89.80 & 72.08 & {\textbf{1.71}} & - & - \\
\cmidrule{2-8}
& JedAI & 95.02 & \textbf{100} & 90.51 & \underline{49.16} & - & - \\
\cmidrule{2-8}
& $\lb$ & 46.09 & \underline{97.10}& 30.21 &  531.63 & \textbf{0.23} & \textbf{0.0039} \\
\cmidrule{2-8}
& $\MS$-1& \textbf{96.21} & 95.36 & \underline{97.07} &  532.17 & 0.45 & \underline{0.32} \\
\cmidrule{2-8}
& $\PM$& \underline{95.15} & 92.85 & \textbf{97.57} & 538.5 & \underline{0.35} & 6.75 \\
\cmidrule{2-8}
& $\ub$ & 91.11 & 85.50& \textbf{97.57} & 531.41 & - & - \\
\midrule
\multirow{6}{*}[-2.5ex]{\rotatebox{90}{\cora}} 
& Magellan &79.70& 93.09 & 69.68 & {\underline{131.13}} & - & - \\
\cmidrule{2-8}
& JedAI & \underline{90.53} & \underline{95.53} & 87.72 & \textbf{40.92} & - & - \\
\cmidrule{2-8}
& $\lb$ &83.57 & \textbf{99.80}& 71.87 &  1,008 & \textbf{7.85} & \textbf{0.29} \\
\cmidrule{2-8}
& $\MS$-1& \textbf{95.55} & 94.25 & \underline{96.79} & 1,031 & 20.88 & \underline{10.50} \\
\cmidrule{2-8}
 & $\PM$&  \textbf{95.55} & 94.25 & \underline{96.79}  & 1,857.6 & 20.19& 837.58 \\
\cmidrule{2-8}
& $\ub$ & 87.50& 79.66 & \textbf{97.05}  & 999.87 & - & - \\
\bottomrule
\end{tabular*}
\caption{Results on Pairwise Matching. Bold figures indicate the best performing results, and the second-best results are underlined.
}
\label{tbl:pwer}
\end{table}

\subsection{Main Results}\label{sec:g-p}
\begin{table*}[htbp]
\renewcommand\arraystretch{0.3}
\setlength{\tabcolsep}{0.25em}
\centering
\begin{tabular*}{\linewidth}{@{}cc|ccc|ccc |cc|ccc|ccc@{}}
\toprule
 \textbf{Data} &  \textbf{Method}  &\multicolumn{3}{c|}{$\fone$~~~~~(\textbf{P}~~  
 /~~~\textbf{R})} &  \textbf{$\ter$} & $\tground$ & $\tsolve$ &\textbf{Data} &  \textbf{Method}  &  \multicolumn{3}{c|}{$\fone$~~~~~(\textbf{P}~~  
 /~~~\textbf{R})} & \textbf{$\ter$} & $\tground$ & $\tsolve$\\

\midrule
\multirow{6}{*}[-2.5ex]{\rotatebox{90}{\imdb}} 
& Magellan & 88.09 & \underline{99.80} & 78.83 & {\textbf{3.89}} & - & - & \multirow{6}{*}[-2.5ex]{\rotatebox{90}{\music}} & Magellan 
 & 89.78& \underline{98.63} & 82.38 & {\textbf{64.83}} & - & - \\
 \cmidrule{2-8} \cmidrule{10-16}
 & JedAI & 97.49& 99.40 & 95.67 & \underline{16.65}  & - & - & & JedAI & 70.67& 87.46 & 59.30 & \underline{100.26} & - & - \\
 \cmidrule{2-8} \cmidrule{10-16}
 &  $\lb$ &72.73& \textbf{100} & 57.15 & 600.65 & \textbf{1.73} & \textbf{0.027} & & $\lb$ &  64.08& \textbf{99.79} & 47.19 & 798.22 & \textbf{25.85} & \textbf{0.071} \\
\cmidrule{2-8} \cmidrule{10-16}
  & $\MS$-1 &\textbf{99.27} &99.39 & \textbf{99.14}  &609.96 & 10.27 & \underline{0.79} & & $\MS$-1 & \textbf{97.52} & 99.25 & \underline{95.58} & 853.91 & 79.75 & \underline{1.86} \\
\cmidrule{2-8} \cmidrule{10-16}
 & $\PM$ & \textbf{99.27} &99.39 & \textbf{99.14} & 643.7 & \underline{9.94} & 34.87 &  & $\PM$ & \textbf{97.52} & 99.25 & \underline{95.58}  & 1,152.4 & \underline{78.64} & 301.51 \\
\cmidrule{2-8} \cmidrule{10-16}
 & $\ub$ &\textbf{99.27} &99.39 & \textbf{99.14} & 598.9 & - & - & & $\ub$ & \underline{97.44}& 99.03 & \textbf{95.90} & 772.3 & - & -\\
\midrule
 \multirow{6}{*}[-2.5ex]{\rotatebox{90}{\cormusic}} & Magellan 
 &  55.54& 97.51 & 38.83 & {66.87} & - & - &\multirow{6}{*}[-2.5ex]{\rotatebox{90}{\poke}} & Magellan & 7.01& 3.97 & 29.74 & {\underline{260.96}} & - & -\\
 \cmidrule{2-8} \cmidrule{10-16}
  & JedAI & 32.75& 73.95 & 21.02 & \textbf{7.88} &- & - & & JedAI & 2.1& 1.08 & 46.56 & \textbf{23.46} & - & - \\
 \cmidrule{2-8} \cmidrule{10-16}
 & $\lb$ & 53.95& \textbf{99.79} &36.97 & 474.56 & \textbf{28.01} & \textbf{0.062} & & $\lb$ & 28.00 & \textbf{100}  & 16.27 & 4,144 & \textbf{2.29} & \textbf{0.018}  \\
 \cmidrule{2-8} \cmidrule{10-16}
 &  $\MS$-1& \textbf{84.10} & \underline{88.11}   & 80.44 & 562.37 & 113.64 & \underline{2.33} & & $\MS$-1&  \textbf{88.71} & \underline{92.88} & \textbf{84.90} & 4,271.8 & \underline{127.67} & \underline{2.17} \\
 \cmidrule{2-8} \cmidrule{10-16}
& $\PM$&  \underline{83.87} & 87.59 & \underline{80.46} & 893.44  & 113.26 & 333.78 & & $\PM$&  \textbf{88.71} & \underline{92.88} & \textbf{84.90} & 4,296 &129.04& 25.84 \\
\cmidrule{2-8} \cmidrule{10-16}
 & $\ub$ & 83.55 & 86.01 & \textbf{81.24} & 446.4 & - & - & & $\ub$ & 77.00 & 70.43 & \textbf{84.90} & 4,142& - & - \\
\bottomrule
\end{tabular*}
\caption{Results on Complex Multi-relational Datasets}
\label{tbl:mrer}
\end{table*}
We evaluated performances of the  lower $\lb$
and upper bound  $\ub$ approximate solutions, a single maximal solution $\MS$-1\footnote{We do not consider all $n$ enumerated maximal solutions as there were only minimal variations among them.},  and all possible merges $\PM$.  We consider $\MS$-1 as the default output of \textsc{ASPen}.
Additionally, we  compared these solutions with two (pairwise) rule-based  ER systems: 
Magellan~\cite{magellan-2016} and JedAI~\cite{jedai-2020}. Note that the two closest approaches, Dedupalog~\cite{ArasuICDE09} and MRL~\cite{DengICDE22}, are not publicly available.
To be comparable, we specified the same preconditions and similarity measures as in~\textsc{ASPen}, and followed the best performing setups for \magellan\footnote{\url{https://tinyurl.com/y6hupmrb}} and \jedai ~\cite{pyjedai-2023}. Note that since both systems lack native support for multi-relational table inputs and do not recognise inter-references between tables, for multi-relational datasets, we performed directly pairwise matching for each table (see Appendix \ref{sec:main-app} for details).

\titlep{Accuracy}~The main results for the pairwise and multi-relational datasets are presented in Tables  \ref{tbl:pwer} and \ref{tbl:mrer}, respectively.
The default output $\MS$-1 consistently achieves the highest F1-score across all datasets, outperforming both \magellan and \jedai by significant margins. On the pairwise benchmarks \dblp and \textit{Cora}, $\MS$-1 surpasses \magellan by 16\% and 1.1\%, and \jedai by 15\% and 5\% respectively. Substantial performance differences are observed in the multi-relational datasets. When comparing $\MS$-1 with \magellan and \jedai, improvements of 11\% and 1.7\%, 7.7\% and 26\%, 28\% and 51\%, and 81\% and 86\%  are observed on \textit{Imdb}, \textit{Mu}, \textit{MuC} and \poke respectively. This shows \ours is promising, particularly the multi-relational setup. 

We also observed that different (approximate) \ours solutions might lead to different performances: 


\titleit{$\ub$ vs $\lb$}
The comparison between $\lb$ and $\ub$ reveals  extreme results in precision and coverage. $\lb$ reached the highest precision (with an average of ${>99}\%$)  in all but one dataset, but has poor coverage with an average recall of $\approx 43\%$. 
Given that duplicated tuples often  contain  different versions of a value, it is not surprising that only few met the strong evidence required by hard rules.
$\ub$ has the best coverage of the considered (approximate) solutions:  on average $\geq 50\%$ higher recall than $\lb$, 0.5\%, 0.26\%, 0.05\%, 0.78\% higher than $\MS$-1 on \dblp, \cora, \music/\cormusic, and slightly higher than $\PM$ on \cora and \music/\cormusic. However, $\ub$ obtains the lowest precision of all solutions  in all datasets. 
This difference 
can be explained by  
the presence of  additional hard rules (the ones replacing soft rules)  for $\ub$, which allow for more merges but also introduce more false positives. 


\titleit{$\MS$-1 \& $\PM$ vs $\ub$ \& $\lb$}
By contrast, the results for 
merge sets 
that make full use of \LACE specifications are more balanced.
On the one hand, $\MS$-1 and $\PM$ outperformed $\lb$ in recall by $\approx 44\%$ on average, with slight sacrifices of precision on most  datasets, only showing a considerable decrease  on \cormusic\ (11\%  lower than $\lb$). 
 On the other hand, $\MS$-1 and $\PM$ outperformed $\ub$ by significant precision margins of 10\% and 7.3\% on \dblp and 14\%  and 22.4\%, on \cora and \textit{Poke}, respectively, with only floating-point drops on recall.
This highlights the effectiveness of combining soft rules and DCs to encourage the discovery of more merges while enforcing consistency to prevent false merges.

\titleit{$\MS$-1 vs $\PM$} 
Observe that $\PM$ and $\MS$-1 have identical results in both precision and recall in \cora, \imdb, \music and \textit{Poke}. A possible explanation is that when DCs are  complementary to the soft rules in a specification, the combination behaves like hard rules~\cite{lace_2022}. With no room for guessing, the specification derives a unique maximal solution, therefore $\MS$-1 and $\PM$ include the same set of merges. Additionally, for \dblp and \cormusic,  we see that precision of $\MS$-1 respectively is 2.5\% and 0.52\% higher than that of $\PM$. For these datasets,  $\PM$ respectively obtained a 0.5\% and 0.05\% higher recall than $\MS$-1. This underscores the characteristic differences between the two type of solutions:
$\MS$-1 prioritises precision while maintaining good coverage, whereas $\PM$ is more inclusive but may consequently contain more false positives.

\titlep{{Running Time}}  
{The overall time, denoted as $\ter$, consists of preprocessing and ER time.   \magellan~and \ours involve preprocessing steps for blocking or similarity filtering, while \jedai~interleaves similarity computation with the ER process, i.e.,\ $\ter$ includes both. 
 For \textsc{ASPen}, the ER time is further composed of  grounding and solving time $\tground$ and $\tsolve$.
 Note that $\ter$ may not always reflect the optimal running times of \textsc{ASPen}. Indeed, when the simplest  $\lb$ setup is considered, directly running an ASP encoding with online similarity evaluation is much faster (cf.\ Appendix  \ref{app:lam}). However, to ensure a consistent analysis of the ER times of \textsc{ASPen}, we present the $\ter$ as the sum of preprocessing and ER time.}

{As shown by the $\ter$ column of Tables~\ref{tbl:pwer} and \ref{tbl:mrer},} both \magellan~and \jedai significantly outperform $\MS$-1  {across all datasets. \magellan~achieved speed advantages of {7.8, 8, 13, 16, 150, 311 times} on \textit{Cora}, \textit{MuC}, \textit{Mu}, \textit{Poke}, \textit{Imdb} and \textit{Dblp} respectively.  
Similarly, \jedai~is  10, 25, 36, 8.5, 71 and 182 times faster across \dblp, \cora, \imdb, \music, \cormusic~and \poke~respectively.
This is largely due to more costly similarity computations in the preprocessing stage, which are necessary to achieve high quality results on the complex multi-relational settings. Indeed, in \music, \cormusic~and \poke, \ours~obtains \emph{substantially higher accuracy} than the baselines.
We compare $\ter$ of other (approximate) solutions with that of {\magellan~and \jedai}~in Appendix~\ref{sec:main-app}.}

When looking at $\tground$ and $\tsolve$ of the different merge sets, $\lb$ is consistently the fastest, with $\tground$ averaging within half a minute and $\tsolve$ concluding in fractions of a second. This performance is expected, given that $\lb$ straightforwardly derives a single set of merges.  
Regarding $\MS$-1,  $\tground$ took the majority of ER time, while solving is consistently much more efficient, terminating in $\leq$ 10 seconds.
For $\PM$,  $\tground$ behaved as for $\MS$-1, but it required longer solving time (100 times longer in the worst case).  This might be explained by the need to perform brave reasoning. We note that  the size of a dataset may not always be the main factor impacting the solving time. For instance, despite containing only 1.9k tuples, $\tsolve$ of $\MS$-1 and $\PM$ on the \cora dataset are noticeably longer than on  much larger datasets  like \music and \textit{Poke}. Similarly, despite \music and \cormusic being of the same size, the $\tground$ of $\ub$, $\MS$-1 and $\PM$ on these datasets are very different. 

\subsection{Effect of Recursion}\label{sec:ab-st}
We executed the levels algorithm described in \sect{sec:asp-alg} on (approximate) solutions derived in our previous experiments on multi-relational datasets. We report accuracy results of $\ub$ and $\MS$-1  on the \poke dataset and merge increments for various recursion levels of \cormusic and \poke in Figures \ref{fig:rec} and \ref{fig:rec-sb}, respectively. Additional results for other datasets can be found in Appendix \ref{sec:recur-app}. We can observe that achieving convergence of merges requires more than one recursion level. For example, in Figure \ref{fig:rec}, \poke converged at level 2 (red dashed line).
\fig{fig:rec-sb} shows noticeable increments of merges obtaining  17.8\% and 18.8\%, 2.4\% and 18.8\% increment gains in the $\MS$-1 and $\ub$ settings on \cormusic and \poke, respectively. This shows the efficacy of recursion in discovering new merges utilising merges derived from previous levels.
\begin{figure}[!htbp]
     \centering
         \centering
         
         \includegraphics[width=0.82\linewidth]{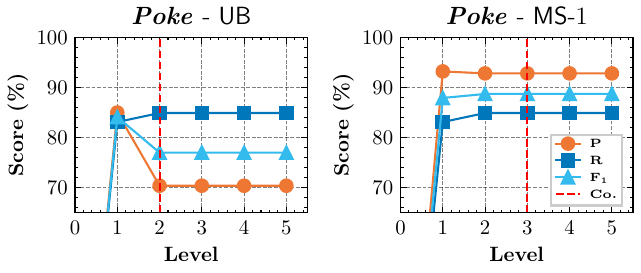}
              \caption{Impact of Recursion on Accuracy} \label{fig:rec}
\end{figure}
\begin{figure}[!htbp]
\includegraphics[width=\linewidth]{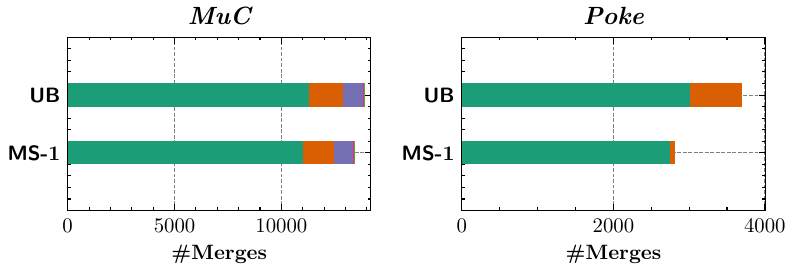}
         \caption{Merge Increments on Levels}
         \label{fig:rec-sb}
     \end{figure}

We present further results on the interplay of recursion  and DCs and dataset characteristics in Appendix~\ref{sec:recur-app}.
\subsection{Efficiency}
\titleit{{Datalog approximations}}
We examined the running times of Datalog programs $\Pi_{\Sigma^\mathsf{lb}}$ and $\Pi_{\Sigma^\mathsf{ub}}$ on all datasets using \ours and the rule engine \vlog~\cite{vlog4j-2019,vlog-2016}. For $\lb$, \ours outperforms \vlog on most datasets, whereas for $\ub$, \vlog is faster on all multi-relational datasets except \textit{Poke}. These results suggest that the performance of different reasoning engines is impacted by characteristics of the data. See Appendix  \ref{sec:eff-app} for more details.

\titleit{{Factors impacting  scalability}}
{Our efficiency analysis shows that both $\tground$ and $\tsolve$ increase monotonically with the \emph{size of the data} and the \emph{percentage of duplicates}. The impact on $\tsolve$ is particularly significant for $\PM$, increasing up to 52 times when these factors are raised by fivefold. Similarly, lowering \emph{similarity thresholds} consistently increases $\tground$, while $\tsolve$ is up to 300 times longer on $\MS$-1 and reaches a time-out ($\geq$24h) on $\PM$. For more details, see Appendix \ref{sec:eff-app}.}

\section{Discussion}~\label{sec:diss}
We have introduced \textsc{ASPen}, an ASP-based implementation of the \textsc{Lace} framework for collective, explainable, and recursive ER. 
Distinguishing features of \ours\ are the consideration of a space of maximal solutions and the ability to supply explanations of derived merges. 
It is also one of only a handful of systems to natively support collective ER tasks, involving multiple database tables and entity types. A comprehensive experimental evaluation 
provided insights into how \textsc{ASPen} performs, in terms of quality metrics and runtime, depending upon which notion of (approximate) solution is employed and how its performance compares to two baseline rule-based ER systems. The overall takeaway is that \textsc{ASPen} is promising, as it is able to terminate in a reasonable amount of time (ER is typically an occasional offline task) and is competitive and often outperforms the baseline systems w.r.t.\ quality metrics, in particular, being able to effectively handle more 
complex multi-relational scenarios. 

Our paper 
showcases entity resolution as an exciting but challenging application for ASP. Indeed, we believe that ER is as an ideal testbed for ASP techniques, as it is an important practical problem that naturally involves many solver functionalities, such as: brave and cautious reasoning (to identify possible and certain merges), preferred answers sets (to generate or reason over maximal solutions), external function calls (for similarity computation), and explanation facilities (to produce justifications of merges). 
By making \textsc{ASPen}'s code and data publicly available, we hope to facilitate future research on ASP-based approaches to ER. 

While our experiments show that \textsc{ASPen} can successfully handle some real-world ER scenarios, scalability remains an issue, and we expect that both general purpose and dedicated optimizations will be needed to be able to scale up to larger datasets and support even more complex reasoning over ER solutions. In addition to continuing to improve the similarity computation phase, we plan to explore the potential of employing specialized data structures or custom procedures for handling equivalence relations, as has been considered for Datalog reasoners \cite{NappaZSS19,SahebolamriBMM23}. As parallelization has been successfully employed in some rule-based ER systems \cite{DengICDE22}, another promising but non-trivial direction would be to see how parallel algorithms can be integrated into \textsc{ASPen}. For this, we hope to build upon existing work on parallelization of Datalog reasoning~\cite{PerriRS13,AjileyeM22} and ASP solving \cite{GebserKS12}.  

We also plan to extend \textsc{ASPen} to handle more expressive ER scenarios, building upon recent extensions to the 
 \textsc{Lace} framework. 
Our top priority will be support not only global merges of entity-referencing constants, as considered in \textsc{ASPen} and the original \textsc{Lace} framework, but also local (cell-level) merges of value constants \cite{DBLP:conf/kr/BienvenuCGI23}, 
so that e.g.\ some occurrences of  ‘J.\ Smith’ can be matched to ‘Joe Smith’ while others are matched to ‘Jane Smith’.
Another important extension, requiring more significant changes to the ASP encoding, would be to allow for both merges and repair operations, as in the \textsc{Replace} framework \cite{DBLP:conf/ijcai/BienvenuCG23}, in order to be able to handle constraint violations that cannot be solved solely via merges. 

\section*{Acknowledgements}
The authors were partially supported by the ANR AI Chair INTENDED (ANR-19-CHIA-0014)
and by MUR under the PNRR project FAIR (PE0000013). 
The authors thank the Potassco community  for their support in resolving the issues encountered while using \clingo.

\bibliographystyle{kr}
\bibliography{kr-sample}

\clearpage
\appendix
\def\consts{\mathbf{C}}



\section{Proof Trees and Explanations}~\label{sec:exp-app}

\begin{figure}[!hbt]
    \footnotesize
    \begin{tikzpicture}[->,thick,level distance=30pt,
        level 1/.style={sibling distance=57pt},
        level 2/.style={sibling distance=60.4pt}]
        \node {$(s_1,s_2)$}
            child {node{$t_1 \approx t_2$}}
            child {node {$\textit{So}(s_1,t_1,W,b_1)$}}
            child {node {$(b_1,b_2)$}
                child {node {$\textit{Ba}(b_1,n_1,g_1,Y,B)$}}
                child {node {$n_1 \approx n_2$}}
                child {node {$\textit{Ba}(b_2,n_2,g_2,Y,B)$}}
                child {node {$g_1 \approx g_2$}}
                }
            child {node{$\textit{So}(s_2,t_2,W,b_2)$}};
    \end{tikzpicture}
    \caption{Proof tree for merge $(s_1,s_2)$ in solution $E_1$. In the tree, $\textit{So}$ and $\textit{Ba}$ stand for the relations $\textit{Song}$ and $\textit{Band}$. $W$, $Y$, and $B$ stand for the constants Waters, 1965, and Barrett, respectively. $n_i$, $g_i$, and $t_i$ are the name and genre of the band $b_i$ and the title of song $s_i$, respectively.}
    \label{ex:ProofTree}
\end{figure}
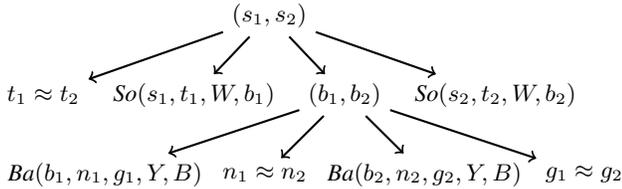

\titleit{Proof Trees} We shall employ proof trees 
to explain 
why a merge $\alpha$ appears in a solution $E \in Sol(D, \Sigma)$.
Formally, we define 
a \emph{proof tree for $\alpha$ in $E$} 
as a node-labelled tree such that (a) the root node has label~$\alpha$, 
(b) every leaf node is labelled with a fact from $D$,
and (c) every non-leaf node~$n$ is labelled with a pair of constants $(d,e)$ 
such that 
one of the following holds:
\begin{itemize}[itemsep=0.5ex, leftmargin=5mm]
\item node $n$ has exactly two children, which have labels $(d,f)$ and $(f,e)$ for some constant $f$ (\emph{transitive node})
\item 
there is a rule\footnote{To avoid an overly lengthy definition, we focus 
on rules without constants, but the definition 
generalises to rules with constants. } 
$P_1(v_1^1, \ldots, v_1^{w_1}) \wedge \ldots \wedge P_m(v_m^1, \ldots, v_m^{w_m})  
 \rightarrow \link(x,y) \in \Gamma$ such that 
 node $n$ has $m$ children labelled with the facts 
$P_1(c_1^1, \ldots, c_1^{w_1}), \ldots, $ $P_m(c_m^1, \ldots, c_m^{w_m}) \in D$, 
there exist $v_i^j=x$ and $v_k^\ell=y$ such that $\{c_i^j, c_k^\ell\} = \{d,e\}$, 
 and additionally, whenever $v_i^j = v_k^\ell$ and $c_i^j \neq c_k^\ell$,
 then $n$ has a child labelled with the pair $(c_i^j, c_k^\ell)$ or $(c_k^\ell, c_i^j)$
(\emph{rule node})
\end{itemize}
Observe that each node corresponds to a single transitive step or rule application, while reflexivity and symmetry steps are left implicit. Also note that when database facts used to satisfy the rule body do not `join', additional children are introduced to ensure that the required merges exist. 
It is not hard to see that every non-trivial merge $\alpha = (c,d) \in E$ (i.e.\ with $c \neq d$) 
has at least one proof tree. A proof tree for the merge $(s_1,s_2)$ in solution $E_1$ is presented in Fig.~\ref{ex:ProofTree}.

We shall also be interested in quantifying the number of successive rule applications 
needed to obtain a given merge. To this end, we define the \emph{rule-depth of a proof tree} $T$
as the maximum number of rule nodes in any leaf-to-root path in $T$. 
The \emph{level of a merge $\alpha$ in a solution $E$} is $0$ if $\alpha=(c,c)$ for some $c\in \consts$,
and otherwise is defined as the minimum rule-depth of all proof trees of $\alpha$ in $E$. 
$(s_1,s_2)$ has level $2$ in solution $E_1$, as it possesses a proof tree with rule-depth $2$ (and no proof tree with rule-depth 1).

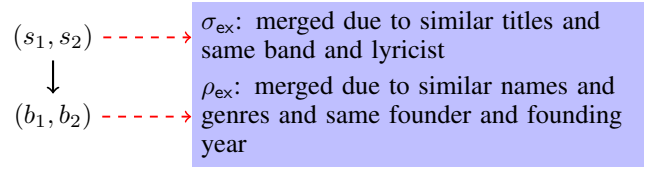
\begin{figure}[!hbt]
    \begin{tikzpicture}[->,thick,node distance=30pt and 138pt,on grid,auto]
        \node (s) {$(s_1,s_2)$};
        \node (b) [below = of s] {$(b_1,b_2)$};
        \node (ds) [right = of s,fill=blue!25,text width=164pt] {$\sigma_\ex$: merged due to similar titles and same band and lyricist};
        \node (db) [right = of b,fill=blue!25,text width=164pt] {$\rho_\ex$: merged due to similar names and genres and same founder and founding year};
        \draw (s) to (b);
        \begin{scope} [->,dashed]
            \draw [line width=0.25mm,  red] (s) to (ds);
            \draw [line width=0.25mm,  red] (b) to (db);
        \end{scope}
    \end{tikzpicture}
    \caption{Example of a graphical proof tree provided by \ours relative to merge $(s_1,s_2)$ in solution $E_1$ for $(D_{\ex},\Sigma_{\ex})$.}
    \label{ex:TreeWithLanguage}
\end{figure}

\titleit{Example Proof Tree from \ours}
An exemplary  proof tree generated by \ours is shown in Figure~\ref{ex:TreeWithLanguage}.
%
\\

     \begin{figure*}[htbp!] 
     \centering
    \begin{subfigure}[b]{0.8\textwidth}
         \centering
         \includegraphics[width=\textwidth]{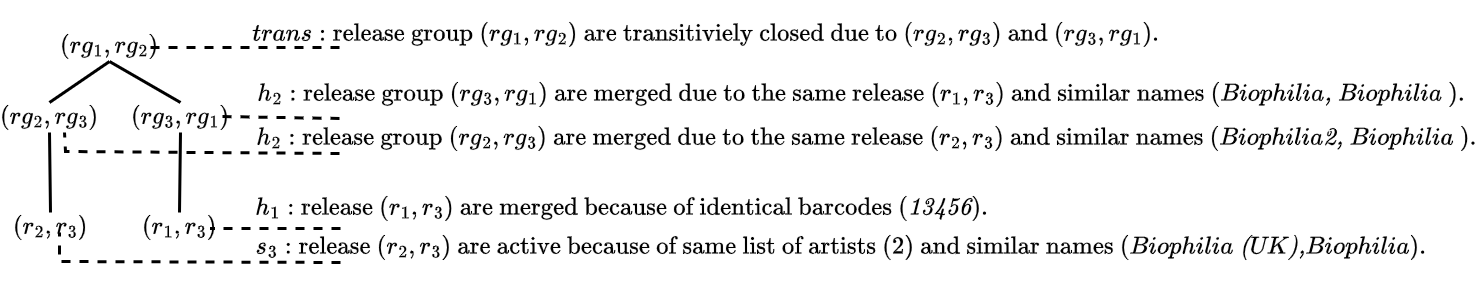}
         \caption{Proof Tree 1 for $(rg_1,rg_2)$ }
         \label{fig:tree1}
     \end{subfigure}
      \hfill
      \begin{subfigure}[b]{0.8\textwidth}
          \centering
         \includegraphics[width=\textwidth]{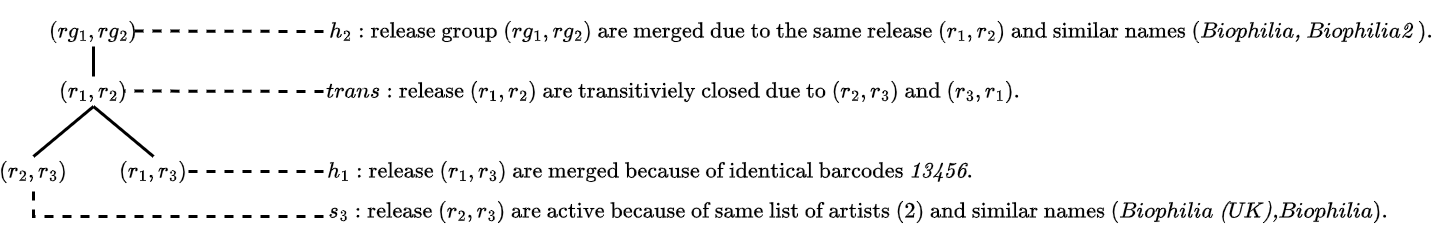}
          \caption{Proof Tree 2 for $(rg_1,rg_2)$}
          \label{fig:tree2}
      \end{subfigure}
      \hfill
              \caption{ Proof trees of  a merge in the \music dataset} \label{fig:trace_tree}
\end{figure*}  
\titleit{{A Qualitative Case}}
In \music, the specification contains a hard rule $h_1$ and a soft rule $s_3$ declaring merges of album releases.  $h_1$: \textit{``two album releases are the same if they have the same barcode"} and $s_3$: \textit{``two album releases are possibly the same if they have similar names and the same list of artists"}. Additionally, it has a hard rule $h_2$ related to release groups: \textit{``two release groups are the same if they have the same release and similar names"}. 

Utilising xclingo~\cite{xclingo}, \ours generates two proof trees for $(rg_1,rg_2)$.
Figure~\ref{fig:trace_tree} illustrates how the merge $(rg_1,rg_2)$ of release group can be justified in a solution. We observe that the solution generated two proof trees for $(rg_1,rg_2)$ as seen in~Figures~\ref{fig:tree1} and \ref{fig:tree2}.

\section{Similarity Computation}\label{app:lam}

\titleit{Loose Upper-bound}
Given an upper-bound transformation $\Sigma^\mathsf{ub}$ of a \textsc{Lace} specification $\Sigma$, let  $\Sigma^{\mathsf{ub}}_{\mathsf{op}}$ be the specification obtained from $\Sigma^\mathsf{ub}$ by dropping every similarity atom in rule bodies. From $\Sigma^\mathsf{ub}_{\mathsf{op}}$, we obtain a Datalog program $P^{\mathsf{ub}}_{\mathsf{op}}$. If we run $P^\mathsf{ub}_{\mathsf{op}}$ over the database $D$, then the extension of $\linkasp$ will contain a loose upper-bound of the pairs of constants that can be potentially merged. An instance of such rule transformed from $\sigma_{\ex}$ in~Figure~\ref{fig:ExLACE} will be:
 \begin{align*} \notag
 \linkasp(X,Y) \leftarrow &\aspsong(X,T,L,B), \aspsong(Y,T^\prime,L,B^\prime),  
        \\ 
        &\defneg~\emptypred(L), \linkasp(B,B^\prime). 
\end{align*}

We executed  the loose upper-bound program for each specification on the datasets on~\vlog~\cite{vlog-2016,vlog4j-2019}. 
Only on \dblp and \cora the programs were able to terminate without throwing errors of \textit{memory overflow}.
We recorded the running times and number of $\linkasp$-facts, comparing with sum of the size of cross products of each merge position pair.

As shown by~\tbl{tbl:loose}, the resulting sets of $\linkasp$-facts for \dblp and \cora barely differ to \textit{\#Cat}. 

\begin{table}[htbp]
            \begin{center}
                \begin{tabular}{cccc}
            \hline 
            \textbf{Data}&  \textbf{\#$\linkasp$}& \textbf{\#Cat}& \textbf{t}(s)\\
            \midrule 
           \dblp & 6,006k& 6,006k& 35.54\\
             \midrule 
             \cora & 3,530k& 3,534k& 23.75\\
              \bottomrule 
                \end{tabular}
 \caption{Size of $\linkasp$-facts in loose upper-bound. \#-columns denote the number of $\linkasp$-facts, the sum of cross-products of constant pairs found in merge positions of a dataset.
            }~\label{tbl:loose}
            \end{center}
        \end{table}  

\titleit{Online vs Preprocessed Similarity}
In principle, one can run directly an ASP program $\Pi_{(D,\Sigma_\mathsf{ext})}$ by replacing literals $\simpred(T,T^\prime, S), S\geq \delta$ with  $\texttt{sim}_{ext}$\texttt{(X,Y)}$\geq\delta$ in $\Sigma$ (and its variants) instead of the method in~\sect{sec:sim}. We conducted a set of experiments comparing the overall running times of \ours when using the approach in \sect{sec:sim} (denoted as $\mathsf{Sim_{opt}}$)
and online similarity evaluation (denoted as $\mathsf{Sim_{ext}}$)
on the datasets and reported the running times as~\tbl{tbl:ol-our}.
%

{We observe that for the simpler solution types $\lb$ and $\ub$, using the external similarity call directly
consistently leads to faster termination times, especially for $\lb$. Indeed, if only $\lb$ or $\ub$ are needed, the similarity algorithm is unnecessary.
For the more complex settings $\MS$-1 and $\PM$, we found that online similarity calculations provided speed advantages on \dblp, \cora, and \cormusic, with improvements of 31\%/32\%, 4\%/3.6\%, and 41\%/25\%, respectively. Conversely, our similarity method performed better on most multi-relational datasets, with improvements of 1.4\%/4.3\%, 27.5\%/22\%, and 16\%/18\% on \imdb, \music, and \poke, respectively. These results suggest that, apart from being able to reuse materialised output ($\oursim$ and the upper-bound merge set $U$) for deriving different solutions, our similarity method can also \textit{speed up} the computation of solutions in complex settings.
}
\begin{table*}[htbp]
\renewcommand\arraystretch{0.3}
\setlength{\tabcolsep}{0.25em}
\centering
\begin{tabular*}{\linewidth}{@{}c|c|cc|cc|c |c|c|cc|cc|c@{}}
\toprule
 \textbf{Data} &  \textbf{Met.}  &  \textbf{$\ter$} &$\tero$ & $\tolground$ & $\tpground$& $\tsolve$ &\textbf{Data} &  \textbf{Met.}  &  \textbf{$\ter$} &$\tero$ & $\tolground$ & $\tpground$& $\tsolve$\\

\midrule
\multirow{2}{*}[-2.5ex]{\dblp  } 
 &  $\lb$ & \rj{531.63}	& \changed{17.12}	& 17.12 &	0.23& 0.0039 &  \multirow{2}{*}[-2.5ex]{\cora} & $\lb$ &  \rj{1,008.01} &	\changed{133.02} &	132.73	&7.85&	0.29\\
\cmidrule{2-7} \cmidrule{9-14}
  & $\MS$-1 & \rj{532.17}	& \changed{362.58}&	362.26	&0.45&	0.32& & $\MS$-1 & \rj{1,031.25} &	\changed{987.75} &	977.25 &	20.88 &	10.5  \\
\cmidrule{2-7} \cmidrule{9-14}
 \multirow{2}{*}{\small{$\tprep= 531$}}  & $\PM$ & \rj{538.5}	& \changed{364.22} &	357.47&	0.35&	6.75&   \multirow{2}{*}{\small{$\tprep = 999$}} & $\PM$ & \rj{1,857.64}	& \changed{1,789.35} &	951.77 &	20.19&	837.58\\
\cmidrule{2-7} \cmidrule{9-14}
 & $\ub$ & \rj{531.4}	& \changed{364.06}	&364.05&	0&	0.0098&  & $\ub$ & \rj{999.87}	& \changed{998.27}	&997.77& 0&	0.5 \\
\midrule
\multirow{2}{*}[-2.5ex]{\imdb} 
 &  $\lb$ & \rj{600.65}	&\changed{78.88}	& 78.86 &	1.73 &	0.027  & \multirow{2}{*}[-2.5ex]{\music} & $\lb$ &  \rj{598.9} & \changed{34.43} & 34.36 & 25.85&	0.071\\
\cmidrule{2-7} \cmidrule{9-14}
  & $\MS$-1 & \changed{609.96}	&\rj{618.62}&	617.83	&10.27	&0.79&  & $\MS$-1 & \changed{853.91}	&\rj{1,178.8}	&1,176.94&	79.75	&1.86 \\
\cmidrule{2-7} \cmidrule{9-14}
\multirow{2}{*}{\small{$\tprep = 598$}}& $\PM$ &\changed{643.71}	&\rj{672.92} &	638.05 &	9.94&	34.87&  \multirow{2}{*}{\small{$\tprep = 772$}}& $\PM$ & \changed{1,152.45}	&\rj{1494.9}&	1,193.39&	78.64&	301.51\\
\cmidrule{2-7} \cmidrule{9-14}
 & $\ub$ &\rj{598.94}	&\changed{535.06}& 535.012 & 0 &	0.045&   & $\ub$ & \rj{772.41}	& \changed{698.91}	&698.8	&0	&0.11 \\
\midrule
\multirow{2}{*}[-2.5ex]{\cormusic} 
 &  $\lb$ & \rj{474.56}&	\changed{29.43}&	29.37&	28.1&	0.062	 & \multirow{2}{*}[-2.5ex]{\poke} & $\lb$ & \rj{4,144.31}	&\changed{7.28}	&7.26&	2.29 &	0.018 \\
\cmidrule{2-7} \cmidrule{9-14}
  & $\MS$-1 &\rj{562.37}	&\changed{331.68}	&329.35&	113.64&	2.33&  & $\MS$-1 &\changed{4,271.86}	&\rj{5,093.16}&	5,090.99	&127.69&	2.17 \\
\cmidrule{2-7} \cmidrule{9-14}
 \multirow{2}{*}{\small{$\tprep = 446$}} & $\PM$ & \rj{893.44}	&\changed{665.53}	&331.7&	113.26	&333.78  &  \multirow{2}{*}{\small{$\tprep = 4,142$}}& $\PM$ &\changed{4,296.8}&	\rj{5,241.05}	&5,215.21&	129.04	&25.84 \\
\cmidrule{2-7} \cmidrule{9-14}
 & $\ub$ &\rj{446.51}	&\changed{306.44}&	306.33&	0	&0.11  & &  $\ub$ & \rj{4,142}	&\changed{4,130.03}&	4,130&	0	&0.035  \\
\bottomrule
\end{tabular*}
\caption{Comparison on solution computation time using  online and preprocessed similarity computation. $\ter$ and $\tero$ record the overall running times of $\mathsf{Sim_{opt}}$ and $\mathsf{Sim_{ext}}$ respectively. $\mathbf{t_p}$ denotes the preprocessing time on a dataset. $\mathbf{t_{og}}$ and $\mathbf{t_{pg}}$ are the grounding time of online and preprocessed similarity respectively, and $\mathbf{t_s}$ is the solving time. Figures in red/green represent worse/better performances between $\ter$ and $\tero$.}
~\label{tbl:ol-our}
\end{table*}

\medskip

\noindent
{\bf Theorem~~\ref{th:simCorrect}}

\titleit{Proof Sketch:}
Let $M^{\textsf{opt}}_\linkasp := \{M_\linkasp \mid M \in \ASet{\Pi_{(D_\mathsf{opt}, \Sigma)}} \}$  and 
$M^{\textsf{all}}_\linkasp  := \{M_\linkasp \mid M \in \ASet{\Pi_{(D_\mathsf{all}, \Sigma)}} \}$.
To show that 
$ M^{\textsf{opt}}_\linkasp  \subseteq M^{\textsf{all}}_\linkasp$, we use the observation that $D_\mathsf{opt} \subseteq D_\mathsf{all}$, and that the rules for encoding hard and soft rules from $\Sigma$ in both programs are the same.
Let $\linkasp(e,e^\prime) \in M^{\textsf{opt}}_\linkasp$, then there is a stable model $M$ of $\Pi_{(D_\mathsf{opt}, \Sigma)}$ containing $\linkasp(e,e^\prime) \in M^{\textsf{opt}}_\linkasp$. Further, there is a rule $\sigma$ in the grounding of  $\Pi_{(D_\mathsf{opt}}$, with $\linkasp(e,e^\prime)$ in the head, with all the atoms $S$ in its body  contained in $M$. Using the observation, we can argue that there is a stable model $M'$ of $\Pi_{(D_\mathsf{all}}$ containing $S$ and therefore  $\linkasp(e,e^\prime) \in M'$, which in turn implies $\linkasp(e,e^\prime) \in M^{\textsf{all}}_\linkasp$.

 To prove that $M^{\textsf{all}}_\linkasp \subseteq  M^{\textsf{opt}}_\linkasp$. Let $\linkasp(e,e^\prime) \in M^{\textsf{opt}}_\linkasp$, and let 
$M' \in \ASet{\Pi_{(D_\mathsf{all}, \Sigma)}}$ a stable model that contains $\linkasp(e,e^\prime)$. Then there is a ground rule $\sigma$ rule in the reduct of $\Pi_{(D_\mathsf{all}, \Sigma)}^M$ and a set of ground atoms $S$ in $M$ that support $\linkasp(e,e^\prime)$.
 If the body of $\sigma$ does not contain an $\linkasp$ atom, then we note that all similarity atoms in the body of $\sigma$ are included in $D_\mathsf{opt}$ by the computation in Phase 3, and because $D_\mathsf{opt}$ occurs in every (stable) model of $\Pi_{(D_\mathsf{opt}, \Sigma)}$ and both programs contain the same rules encoding $\Sigma$, is easy to see that $\linkasp(e,e^\prime)$ occurs in a stable model of  $\Pi_{(D_\mathsf{opt}, \Sigma)}$. For the case where $\sigma$ contains $\linkasp$ atoms, we can use an inductive argument, with the previous case being the base of the induction. An important observation to construct the argument is that all the similarity atoms used in the derivation of $\linkasp(e,e^\prime)$ w.r.t  $\Pi_{(D_\mathsf{all}, \Sigma)}^M$  are added to $\Pi_{(D_\mathsf{opt}, \Sigma)}$ either in Phase 1 or in Phase 3 of the similarity computation.

\section{Experimental Setup}~\label{sec:setup-app}



\begin{table}[!htp]
\renewcommand\arraystretch{0.8}
\setlength{\tabcolsep}{1em}
\centering
\begin{tabular*}{\linewidth}{@{}cccccc@{}}
\toprule
\textbf{Name}& \textbf{\#Rec}& \textbf{\#Rel}& \textbf{\#At} &\textbf{\#Ref} &\textbf{\#Dup}\\
\midrule
\dblp & 5k & 2 & $5^{b}$ & 0 & 2.2k\\
     \midrule
\cora & 1.9k & 1 & 17 & 0 & 64k\\
    \midrule
\imdb& 30k & 5& 22 & 4 & 6k \\
       \midrule
\music& 41k & 11& 72 & 12 & 15k \\
    \midrule
\cormusic& 41k & 11& 72 & 12 & 15k\\
      \midrule
\poke & 240k & 20& 104 & 20 & 4k \\
\bottomrule
\end{tabular*}
\caption{Dataset Statistics.  \#-columns represent the number of records, relations, attributes, referential constraints and duplicates, respectively. $\cdot^{ b}$: the DBLP and ACM tables share the 5 attributes.
}\label{tbl:ds-stats}
\end{table}

\titleit{Datasets} Statistics of the datasets used in our experiments are shown in  Table~\ref{tbl:ds-stats}. Note that regardless of the number of relations and attributes, datasets with a larger number of referential constraints are  structurally more complex. Sources of the original databases can be found at:
\enumi~\cora: \url{https://hpi.de/naumann/projects/repeatability/datasets/cora-dataset.html},
\enumii~\music: \url{https://musicbrainz.org/doc/MusicBrainz\_Database/Schema}, \enumiii~\poke: \url{https://pokemondb.net/about}.

\titleit{Similarity Measures and Metrics} 
We calculate the syntactic similarities of constants based on their data types~\cite{erblox-2017}.   \textbf{(i)} For numerical constants, we use the Levenshtein distance; \textbf{(ii)} for short string constants (length$ <25$), we compute the score  as the editing distance of two character sequences (Jaro-Winkler distance);  \textbf{(iii)} for long-textual constants (length$\geq 25$), we use the TF-IDF cosine score as the syntactic similarity measure.  We adopt  the commonly used metrics~\cite{DBLP-10} of \emph{Precision} (\textbf{P}), \emph{Recall} (\textbf{R}) and \emph{F1-Score} ($\mathbf{F_1}$) to examine the solutions derived from the specifications. Precision reflects the percentage of true merges in a solution and Recall indicates the coverage of true merges in a solution relative  to the ground truth. The quality of a solution is then measured as $\text{F}1 = 2\times\text{Precision}\times\text{Recall}/(\text{Precision}+\text{Recall})$.

\titleit{Clean Data Sampling} {We describe the process of data sampling and synthesising duplication for \music/\cormusic and \poke datasets.}
{Note that it is important to preserve the relation between entities from different relations when corrupting the instances to retain interdependencies of the data. Thus, we consider the referential dependency graphs of the schema when creating the datasets. Assuming the original schema instances of the \music and \poke are clean, we sampled tuples from each table and created clean partitions of the instances. In particular, we started from the relations with zero in-degree and sampled for each step the adjacent referenced entity relations that have all their referencing relations sampled. Since tuples from relations store only foreign keys do not represent entities,
they were sampled only after one of their referencing relations were already sampled. }

\begin{figure}
    \centering
    \includegraphics[width=0.5\textwidth]{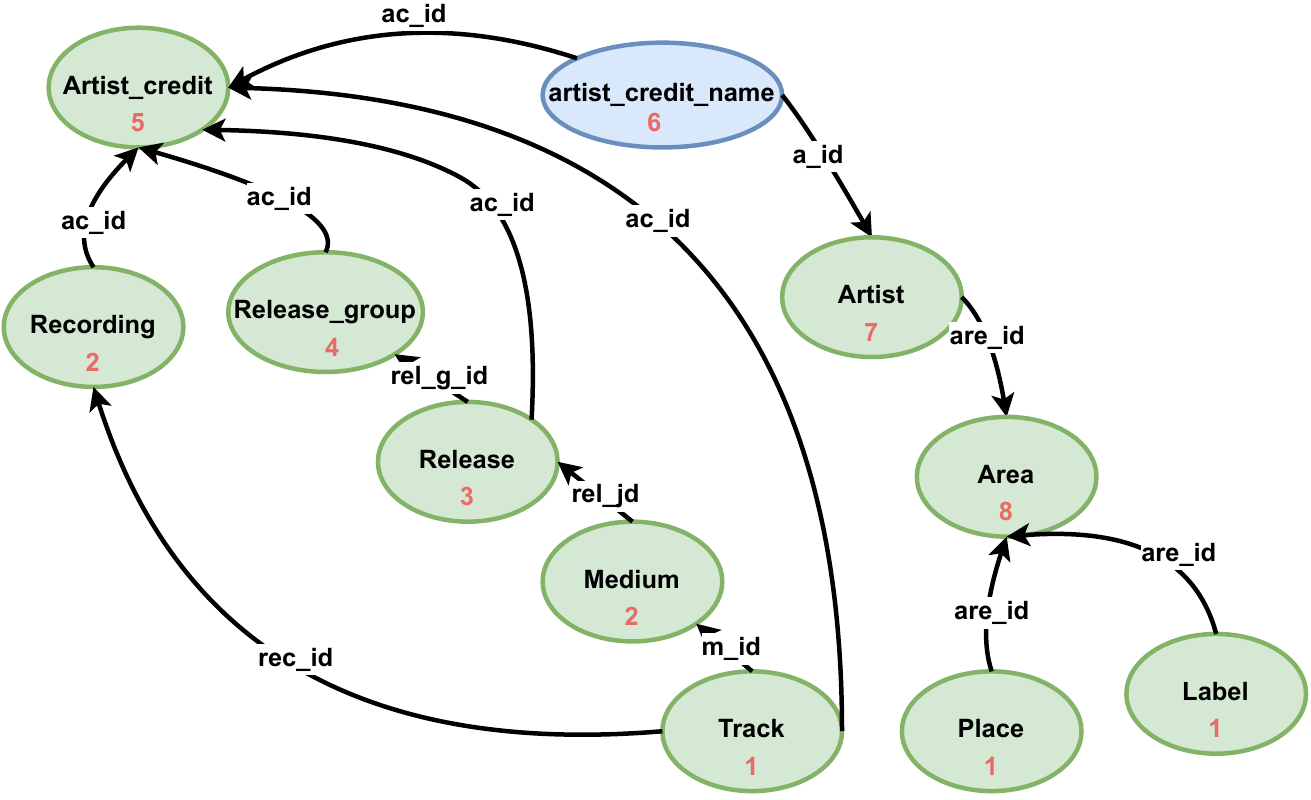}
    \caption{The relational dependencies of \music }\label{fig:music-dp}
\end{figure}

{In Figure~\ref{fig:music-dp}, green nodes and blue nodes respectively represent entity and non-entity relations in the \music schema and edges represent  referential constraints from an out-node to an in-node labelled with the corresponding foreign key. In this case, we begin sampling from the entity relations \texttt{Track}, \texttt{Place}, \texttt{Label} since they are not referenced by any other relations. As \texttt{Artist\_Credit} relation is also referenced by many other relations, in the second step we sample only those that are adjacent to \texttt{Track} and with all in-arrows sampled, i.e., the \texttt{Medium} and \texttt{Recording} relations. Entities of other relations are sampled analogously. Note that since the non-entity relation \texttt{Artist\_Credit\_Name} stores mappings between  \texttt{Artist\_Credit} and \texttt{Artist}, it is sampled only after the entities of \texttt{Artist\_Credit} are picked. We are then able to proceed sampling from \texttt{Artist} when \texttt{Artist\_Credit\_Name} is selected. Consequently, clean partitions of the schema instances can be obtained from sampling. The sampling procedure is done by an extra ASP program as a part of data preprocessing step.}

\titleit{Duplicates Generation}  The original \music and \poke datasets are clean, so we synthesised and injected duplicates to create the datasets with duplicates employed in our experiments. To this end, we utilised the Geco corruptor~\cite{geco-13}. The corruptor randomly picked entities from the clean instances and generated per record up to 3 duplicates. 
Errors of different types, such as keyboard input errors, OCR (characters visually similar) errors, and null values were injected following a predefined distribution across tuple attributes to ensure uniqueness of the  duplicates created. 
Importantly, tuples may contain foreign keys, so it is undesirable that the generated duplicates are not referenced by other tuples at all. Hence, for each tuple with foreign keys, we replaced each foreign key $k$ (which we always consider as merge attributes) with an identifier randomly drawn from the equivalence class of $k$ (including $k$ and its duplicates) as a type of error injection after creating duplicates.  Finally, the ground truth of the datasets is obtained from the identifiers of the original clean instances and their duplicates.

\titleit{Environment}
We implemented \ours in Python. The ER controller, \texttt{asprin} optimiser~\cite{asprin-2015}, and explainer~\texttt{xclingo2}~\cite{xclingo2} are all based on \texttt{clingo 5.5}~\footnote{https://potassco.org/clingo/python-api/5.5/} Python API. The specification of programs follow the format of the  ASPCore2.0 standard~\cite{asp-core-2022}. All the experiments were run on a workstation using a single 3.8GHz  AMD Ryzen Threadripper 5965WX core and 128 GB of RAM.


    

\section{Main Results}~\label{sec:main-app}
\titleit{Multi-relational Input to Baselines}
Note that since baseline systems lack native support for multi-relational table inputs and do not recognise inter-references between tables, for multi-relational datasets. Moreover, these approaches assume that only one merge position is present for each relation (hence consider tuples are entities). Therefore, we performed directly pairwise matching for each table. Specifically, let $\mathcal{S} = \{R_1,...,R_n\}$ be a multi-relational schema  and $D$ be a $\mathcal{S}$-database, we consider pairwise matchers a function $m:D\times D \rightarrow \{0,1\} $, where  0 and 1 represent not match/match resp. The set of merges w.r.t. $D$ is collected as 
\begin{equation*}
    \{ (t_j,t^\prime_j) \mid m(R_i(t_1,...,t_k),R_i(t^\prime_1,...,t^\prime_k))=1, R_i\in \mathcal{S}\}
\end{equation*}
where $j$ is a merge position. 
\\
{\titleit{Running Time}
The preprocessing time is used as $\ter$ for $\ub$, as $\ub$ can be obtained directly from this step. When comparing baselines with various solutions of \ours, it is evident that all \ours (approximate) solutions are significantly slower than the baselines. This is primarily due to the costly preprocessing stage. The $\lb$ and $\ub$ times are comparable, being 310 and 10, 7.6 and 24.6, 154 and 36, 12 and 7.9, 7.1 and 60.2, 15.8 and 176 times slower than \magellan and \jedai on \dblp, \cora, \imdb, \music, \cormusic, and \poke, respectively. The worst performance is observed on the most complex setup $\PM$, where \ours is up to 340 and 183 times slower than \magellan and \jedai, respectively.}
\smallskip 

\titleit{Dirtiness of Duplicates} 
We observe that in \imdb and \music, 
$\ub$, $\MS$-1 and $\PM$ achieved nearly perfect F1 scores. Remarkably,  results are identical for the three type of solutions in \imdb.
This uniformity may indicate that values on duplicates of entities exhibit low variance, resulting in a `cleaner' instance. Indeed, if values of duplicates of an entity are largely identical, the discovery of merges becomes easier as merges derived from soft rules  become more certain.
Clearly, if merges derived from soft rules are as certain as those raised from hard rules, DCs would not be triggered at all.  
This observation is confirmed by the  differences  in accuracy between $\MS$-1 and $\PM$  on the dirtier \cormusic.  Although the number of duplicates per entity in \cormusic is the same as in \music, the presence of more variants and nulls in \cormusic may have introduced more uncertainties.

\section{Effect of Recursion}~\label{sec:recur-app}

We executed the levels algorithm described in \sect{sec:asp-alg} on solutions derived in our previous experiments on multi-relational datasets. 

\begin{figure*}[htbp] 
     \centering
     \begin{subfigure}[b]{0.8\textwidth}
         \centering
        \includegraphics[width=\textwidth]{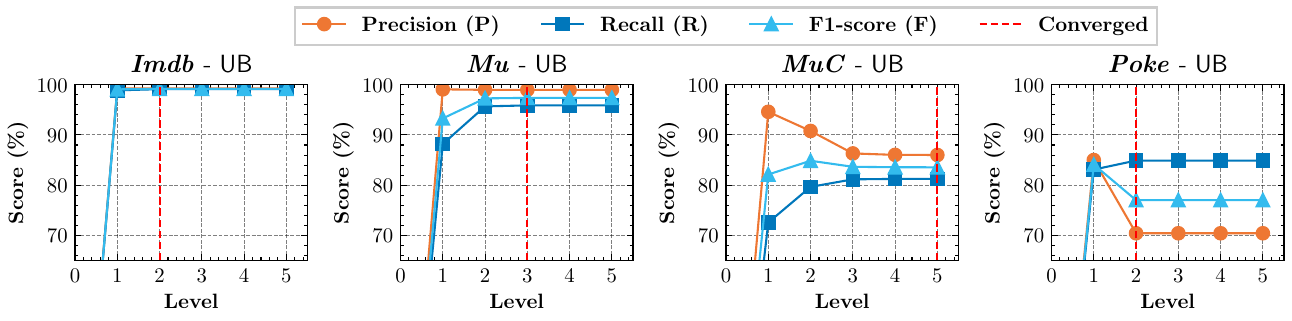}
         \caption{Multi-level recursion on $\ub$ 
         }
         \label{fig:rec-ub}
     \end{subfigure}
     \hfill
     \begin{subfigure}[b]{0.8\textwidth}
         \centering
         \includegraphics[width=\textwidth]{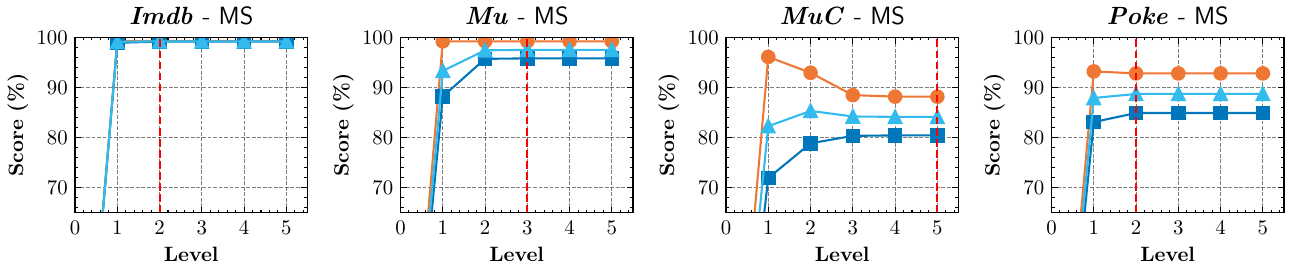}
         \caption{Multi-level recursion on $\MS$-1 }
         \label{fig:rec-ms}
     \end{subfigure}
     \hfill
       \begin{subfigure}[b]{\textwidth}
       \includegraphics[height=0.17\textwidth,width=\textwidth]{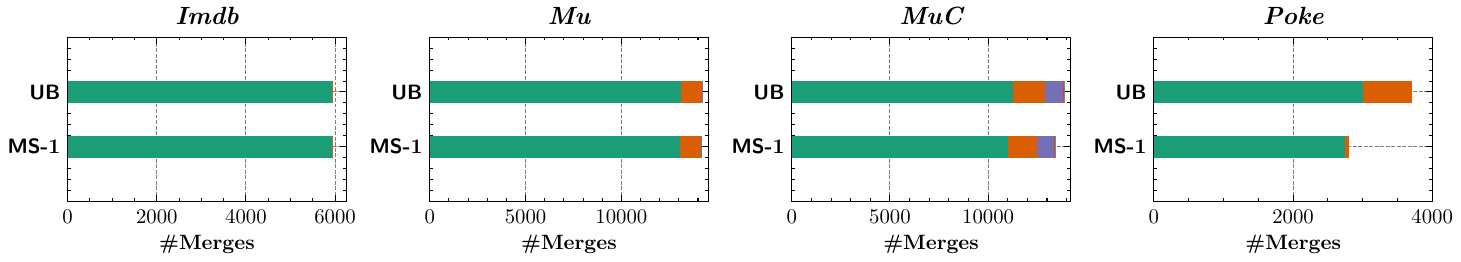}
         \caption{Merge Increments on Levels}
         \label{fig:rec-sb2}
     \end{subfigure}
       \hfill
              \caption{Effects of Recursion} \label{fig:rec2}
\end{figure*}

\titleit{Recursion is Effective} In general, achieving convergence of merges requires more than one recursion level:  \imdb and \poke converged at level 2, while \music and \cormusic converged at level 3 and level 5, respectively.
As illustrated by \fig{fig:rec-sb2}, noticeable increments of merges are observed in most of the datasets, obtaining 8\% and 8.03\%, 17.8\% and 18.8\%, 2.4\% and 18.8\% increment gains in $\MS$-1 and $\ub$ settings on \music, \cormusic and \poke respectively. These patterns confirm the efficacy of recursion in discovering new merges utilising merges derived from previous levels.

\titleit{DCs are Important for Recursion}
The impact of DCs on recursion becomes apparent when comparing the performances depicted in \fig{fig:rec-ub} and \fig{fig:rec-ms} . While recursion generally improves F1 scores across most datasets, it is interesting to observe that in \poke, introducing recursion leads to a decline in accuracy of $\ub$. Notably, at the second level of $\ub$, a slight increase in recall comes at the expense of a significant drop in precision, resulting in an 8\% decrease in F1 score. Conversely, with the integration of DCs, the precision of $\MS$-1 on \poke remains consistent,  ultimately leading to an increase in F1 score. This highlights the critical role of DCs in recursive setups, enhancing the consistency of newly included merges across subsequent levels.

\titleit{Recursion and Dataset Characteristics}
 One can observe in \fig{fig:rec-sb2}  that \imdb shows only a slight increase in the number of merges, whereas \music, \cormusic, and \poke (despite a small increment for $\MS$-1 due to the regulation of DCs) exhibit more substantial increments in levels $>1$. This disparity in merge counts correlates with the increasing number of references within the databases.  Indeed, as all  merge attributes specified are key attributes,   databases with a greater number of inter-table references are better poised to leverage recursion for merge identification.
When comparing \music and \cormusic, it is interesting to observe that despite having the same number of referential constraints, \cormusic yields 2,600 more merges in later levels and requires two additional iterations to converge. This parallels the findings in~\sect{sec:g-p}, suggesting that recursion may prove more effective in discovering merges when dealing with dirtier duplicates. Indeed, simply comparing the  similarity of attributes is less likely to identify merges due to the presence of nulls and value variations. Consequently, identifying duplicates may necessitate to consider the inter-dependencies between entities within the dataset.

\section{Efficiency}\label{sec:eff-app}
We conducted two sets of experiments to examine the 
efficiency of Datalog approximations and impact on efficiency of various factors: \enumi~data size; \enumii~the percentage of duplicates  in a dataset; \enumiii~similarity thresholds on an ER program.

\titlep{Datalog Approx.}
We assume similarity facts are given by $\oursim$ and ran the Datalog programs  $\Pi_{\Sigma^\mathsf{lb}}$ and $\Pi_{\Sigma^\mathsf{ub}}$ on all datasets using \ours and \vlog~\cite{vlog4j-2019,vlog-2016}.
\tbl{exp:datalog} presents the running time results. For $\lb$, \ours outperforms \vlog on most datasets, being 1.03, 1.7, 6 and 9 times faster on \textit{Imdb}, \textit{Cora}, \textit{Poke} and \textit{Dblp}, respectively. Conversely, \vlog terminated 4 and 11 times faster on \music and \cormusic. For $\ub$, \vlog outspeeds \ours in all but one of the multi-relational datasets, being 4.3, 7 and 46 times faster on \textit{MuC}, \textit{Imdb} and \textit{Mu}, respectively. 
However, \ours surpassed \vlog by 7 times on \poke. This results suggest that the performance of different reasoning engines is impacted by characteristics of the data.
\begin{table}[!htp]
\renewcommand\arraystretch{0.3}
\setlength{\tabcolsep}{0.24em}
\centering
\begin{tabular*}{\linewidth}{@{}cccccccc@{}}
\toprule
\textbf{Prg.} & \textbf{Sys.}  &$\mathbf{t_\dblp}$ & $\mathbf{t_\cora}$ & $\mathbf{t_\imdb}$ & $\mathbf{t_\music}$ &$\mathbf{t_\cormusic}$ & $\mathbf{t_\poke}$\\
\midrule
\multirow{2}{*}{$\lb$} & \vlog&  0.76 & 8.06 &          6.02 & \textbf{2.33} & \textbf{7.25} & 39.86\\
\cmidrule{2-8}
 &            \ours &  \textbf{0.079} & \textbf{4.66} & \textbf{5.8} & 27.03 & 30.21 & \textbf{6.38}\\
\midrule
\multirow{2}{*}{$\ub$} &\vlog& 0.83 & 10.8 &  \textbf{32.19}& \textbf{6.66} & \textbf{15.47} & 5115.31\\
\cmidrule{2-8}
 &  \ours                    &  \textbf{0.13} & \textbf{7.74} & 240.51 & 307.53  & 67.63 & \textbf{697.96}\\
\bottomrule
\end{tabular*}
\caption{VLog4j vs. \ours on $\lb$ and $\ub$.}
\label{exp:datalog}
\end{table}

\subsubsection{Varying Size of the Data}~\label{sec:vds}
We ran \ours on variants of \music, where the data size $|D|$ ranged from $\times1$ to $\times5$,  maintaining a consistent 10\% proportion of duplicates (higher than the real-world duplicate distribution of 1\%~\cite{er-data-ijcai-2022}). \tbl{exp:vds} illustrates the changes in grounding and solving times on both $\MS$-1 and $\PM$. Overall, both $\tground$ and $\tsolve$ increase monotonically as $|D|$ increases. $\tground$ follows a similar pattern for both $\MS$-1 and $\PM$, increasing by factors of 5, 22, 42, and 63 as the data scale increases from $\times2$ to $\times5$. However, while $\tsolve$ increases linearly for $\MS$-1, it increases more drastically for $\PM$, resulting in running times 6, 16, 29, and 52 times longer across the range of sizes $|D|$.
\begin{table}[!htp]
\renewcommand\arraystretch{0.3}
\setlength{\tabcolsep}{0.7em}
\centering
\begin{tabular*}{\linewidth}{@{}ccccccc@{}}
\toprule
\textbf{$\mathbf{|D|}$}  & \textbf{Met.} & $\tground$ & $\tsolve$ &\textbf{Met.} & $\tground$ & $\tsolve$ \\
\midrule
 $\times 1$&  \multirow{5}{*}[-2.0ex]{\rotatebox{90}{$\MS$-1}} & 17.82 & 1.02 &   \multirow{5}{*}[-2.5ex]{\rotatebox{90}{$\PM$}}&  17.55	& 14.87 \\ 
\cmidrule{3-4} \cmidrule{6-7}
$\times 2$&  &  92.4	&2.21	&   &   92.75	& 84.6	 \\ 
\cmidrule{3-4} \cmidrule{6-7}
$\times 3$&  &  388.2	&3.4	 &   &418.17	&228.87	\\ 
\cmidrule{3-4} \cmidrule{6-7}
$\times 4$&  &  719.2	&4.8	  &   &763.61	&416.48\\ 
\cmidrule{3-4} \cmidrule{6-7}
$\times 5$&  &  1083.6	&5.7	&   &1106.7	&735.43\\ 
\bottomrule
\end{tabular*}
\caption{Impact of Varying $|D|$}
\label{exp:vds}
\end{table}

\subsubsection{Varying  the Percentage of Duplicates}~\label{sec:vdup}\par
\titleit{Setup} We created variants of \music with  duplication ratios of $10\%$, $30\%$ and $50\%$, while keeping the same dataset size. We denote the variants  as $\dup\{10,30,50\}$, respectively. To mitigate the potential impact brought by tuple distribution, the sets of duplicates are such that $\dup{10}$ $\subset$ $\dup{30}$$ \subset$ $\dup{50}$. 

 \titleit{Result} 
  The results are presented in~\fig{fig:var-dup}. Since $\lb$, $\ub$ and $\MS$-1 present a similar behaviour, we only include the results of $\MS$-1 and $\PM$ for comparison.  Generally, increasing the proportion of duplicates led to monotonic increases in both $\tground$ and $\tsolve$. This explains the previous observation of longer times spent on smaller datasets like \cora. Indeed, despite its smaller scale, the ground truth size of \cora is much larger than that of other datasets. The increase in solving time $\tsolve$ for $\PM$ in $\dup{50}$ was particularly notable, being almost 30 times longer than in $\dup{10}$. This can be attributed to the larger number of possible merges derived from soft rules due to the increased size of duplicates, leading to a larger space of merge combinations and consequently longer solving times.
\subsubsection{Varying Similarity Thresholds}~\label{sec:vsim}\par
\begin{figure}[htbp]
    \centering
    \begin{subfigure}[b]{0.45\linewidth}
        \centering
        \includegraphics[width=\linewidth]{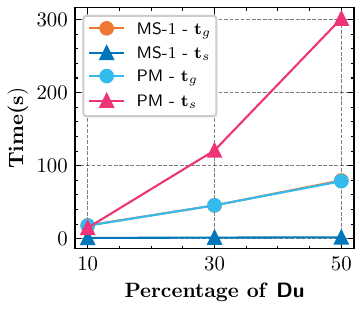}
        \caption{Varying $\dup$}
        \label{fig:var-dup}
    \end{subfigure}
    \hfill
    \begin{subfigure}[b]{0.52\linewidth}
        \centering
        \includegraphics[width=\linewidth]{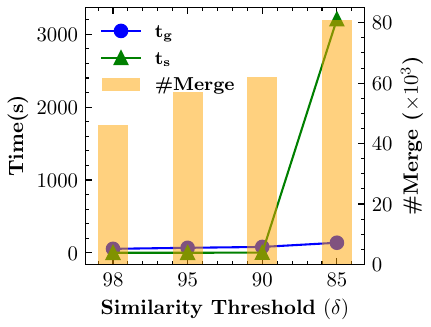}
        \caption{Varying $\delta$}
        \label{fig:sim-eff}
    \end{subfigure}
    \caption{Impacts on duplicate (\%) and similarity threshold}
    \label{fig:overall}
\end{figure} 
\titleit{Setup} We create variants of the ER program for \music by adjusting similarity thresholds.  This results in four versions of the \music specification with similarity thresholds set to 98, 95, 90, and 85. For simplicity, we refer to the similarity threshold as $\delta$.


\titleit{Results}
We ran the ER program with different $\delta$s across the solution setups on \music. 
We observed consistent patterns in $\tground$ across all setups  ($\lb$, $\ub$, $\MS$-1, and $\PM$) with an approximately nine-fold increase in $\tground$ when reducing $\delta$ from 98 to 85, as exemplified  for $\MS$-1  in~\fig{fig:sim-eff}.
This increase can be attributed to the nature of the recursive evaluation algorithm, e.g.  semi-naive evaluation~\cite{asp-in-prac-2012}, adopted by the grounder.
As the value of $\delta$ lowers, more new merges may  be produced in earlier levels, so that they can be reevaluated in  later ones. This is shown  in  the trend of orange charts w.r.t. merge increments. 
A more interesting pattern is observed when looking at the solving times on $\MS$-1 and $\PM$.  We can observe a dramatic increase of $\tsolve$ when reducing $\delta$ from 90 to 85: as shown by the green curve in~\fig{fig:sim-eff}, $\tsolve$ for $\MS$-1 sharply rose from tens of seconds to 3,345 seconds. Moreover, $t_s$ reached a timeout ($> 24$ hours) for $\PM$ with $\delta = 85$. This suggests the scalability of \ours may potentially be restricted in complex reasoning settings, provided the inherent intractability of computing $\MS$ and $\PM$~\cite{lace_2022}.

\end{document}